\documentclass[a4paper,fleqn]{cas-dc}

\usepackage[numbers]{natbib}
\usepackage{amsmath}
\usepackage{color}
\usepackage{algorithmic}
\usepackage[percent]{overpic}
\usepackage{array}

\usepackage{fixltx2e}
\usepackage{stfloats}
\usepackage{float}
\usepackage{textcomp}
\usepackage{multirow}

\usepackage[font=small,labelfont=bf,labelsep=none]{caption}

\def\etal{\emph{et al.}}
\def\ie{\emph{i.e.}}
\def\eg{\emph{e.g.}}

\def\tsc#1{\csdef{#1}{\textsc{\lowercase{#1}}\xspace}}
\tsc{WGM}
\tsc{QE}
\tsc{EP}
\tsc{PMS}
\tsc{BEC}
\tsc{DE}

\begin{document}
\let\WriteBookmarks\relax
\def\floatpagepagefraction{1}
\def\textpagefraction{.001}
\shorttitle{Segment Anything Model for Medical Image Segmentation: Current Applications and Future Directions}
\shortauthors{Yichi Zhang et~al.}

\title [mode = title]{Segment Anything Model for Medical Image Segmentation: Current Applications and Future Directions}

\author[1]{Yichi Zhang}
\author[2]{Zhenrong Shen}
\author[2]{Rushi Jiao}

\address[1]{School of Data Science, Fudan University, Shanghai, China}
\address[2]{School of Biomedical Engineering, Shanghai Jiao Tong University, Shanghai, China}

\begin{abstract}
Due to the inherent flexibility of prompting, foundation models have emerged as the predominant force in the fields of natural language processing and computer vision.
The recent introduction of the Segment Anything Model (SAM) signifies a noteworthy expansion of the prompt-driven paradigm into the domain of image segmentation, thereby introducing a plethora of previously unexplored capabilities. 
However, the viability of its application to medical image segmentation remains uncertain, given the substantial distinctions between natural and medical images.
In this work, we provide a comprehensive overview of recent endeavors aimed at extending the efficacy of SAM to medical image segmentation tasks, encompassing both empirical benchmarking and methodological adaptations.
Additionally, we explore potential avenues for future research directions in SAM's role within medical image segmentation.
While direct application of SAM to medical image segmentation does not yield satisfactory performance on multi-modal and multi-target medical datasets so far, numerous insights gleaned from these efforts serve as valuable guidance for shaping the trajectory of foundational models in the realm of medical image analysis.
To support ongoing research endeavors, we maintain an active repository that contains an up-to-date paper list and a succinct summary of open-source projects at \url{https://github.com/YichiZhang98/SAM4MIS}.
\end{abstract}




\maketitle

\section{Introduction}
Medical imaging is at the forefront of healthcare, playing a crucial role in the diagnosis and treatment of diseases.
Medical image segmentation aims to discern specific anatomical structures from medical images including organs, lesions, tissues, etc.
This fundamental step is integral to numerous clinical applications, such as computer-aided diagnosis, treatment planning, and monitoring disease progression~\cite{MIA2017survey,zhou2021review}. 
Accurate segmentation can provide reliable volumetric and shape information of target structures, thereby assisting in many further clinical applications like disease diagnosis, quantitative analysis, and surgical planning \cite{AbdomenCT-1K,medicaldecathlon,totalsegmentator}.
Deep learning models have demonstrated great promise in the field of medical image segmentation due to their ability to learn intricate imaging features. 
However, existing methods are often tailored for specific modalities or targets, which constrains their ability to effectively generalize across different tasks.

The emergence of large-scale foundation models~\cite{wang2023large,liang2022foundations} has revolutionized artificial intelligence and sparked a new era due to their remarkable zero-shot and few-shot generalization abilities across a wide range of downstream tasks \cite{awais2023foundational}. It is of great importance to develop foundation models capable of adapting to various imaging modalities to tackle complex segmentation tasks for medical imaging \cite{ma2023towards}.
With the recent introduction of Segment Anything Model (SAM)~\cite{SAM-Meta} as a pioneering foundational model for image segmentation, it has gained massive attention due to its strong capabilities for generating accurate object masks in a fully automatic or interactive way. 
This introduction marks the inception of the prompt-driven paradigm into the domain of image segmentation, unlocking a myriad of previously unexplored capabilities.
However, as a very important branch of image segmentation, it remains unclear whether these foundation models can be applied to medical image segmentation due to the essential differences between natural images and medical images.
To this end, a large number of extended works have been proposed by the community to further explore the usage of SAM for medical image segmentation.

In this paper, we aim to summarize recent efforts to extend the success of SAM to medical image segmentation tasks.
Firstly, we briefly introduce the background of foundation models and the workflow of SAM. 
After that, we review and divide current works into two main directions.
The first direction aims to evaluate the zero-shot performance of SAM with different prompt modes in various medical image segmentation tasks, while the other one focuses on exploring methods for SAM adaptation to medical image segmentation.
Finally, we conclude the survey and outline several existing challenges as well as potential future directions.
To keep up with the rapid increase of the research community, we maintain a continuously updated paper list and open-source project summary to boost the research on this topic.

\begin{figure*}
        \centering
	\begin{overpic}[width=18cm]{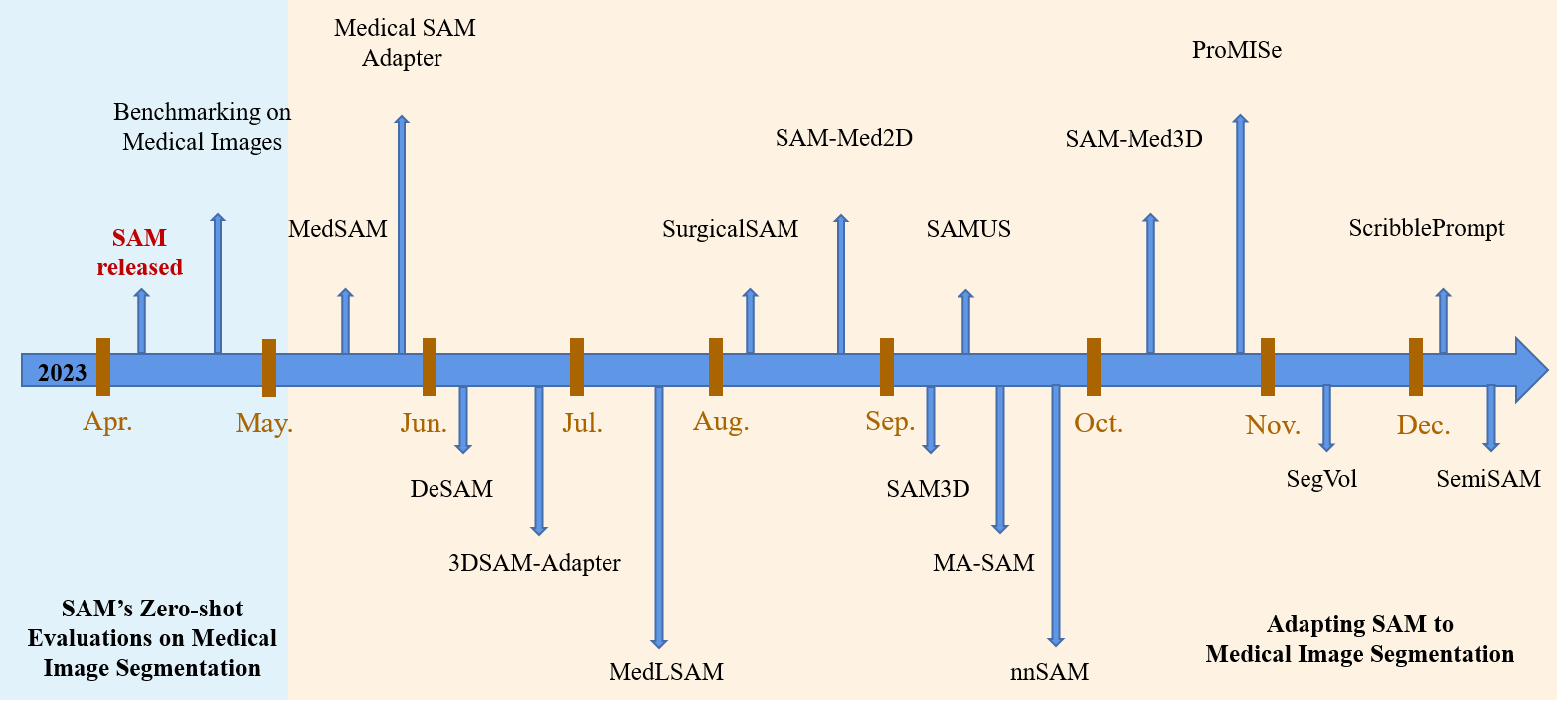}
        \put(10.7,29){\cite{SAM-Meta}}
        \put(11,33){\cite{SAM-Empirical,SAM-SZU}}
        \put(21,27.5){\cite{MedSAM}}
        \put(25,38.5){\cite{Med-SAM-Adapter}}
        \put(27.5,11){\cite{DeSAM} }
        \put(33,6){\cite{3DSAM-adapter} }
        \put(47,1.5){\cite{MedLSAM}}
        \put(46,28){\cite{SurgicalSAM} }
        \put(52,33.5){\cite{SAM-Med2D}  }
        \put(58,11){\cite{bui2023sam3d}  }
        \put(60.5,28){\cite{lin2023samus} }
        \put(62,6){\cite{MA-SAM} }
        \put(70,1.5){\cite{li2023nnsam} }
        \put(71.5,33.5){\cite{SAM-Med3D} }
        \put(78,39.2){\cite{li2023promise} }
        \put(82.5,11.5){~\cite{du2023segvol} }
        \put(90,28){\cite{scribbleprompt} }
        \put(94,11.5){\cite{zhang2023semisam} }
        \end{overpic}
	\caption{A brief chronology of Segment Anything Model (SAM)~\cite{SAM-Meta} and its variants for medical image segmentation in 2023. }
	\label{trends}
\end{figure*}
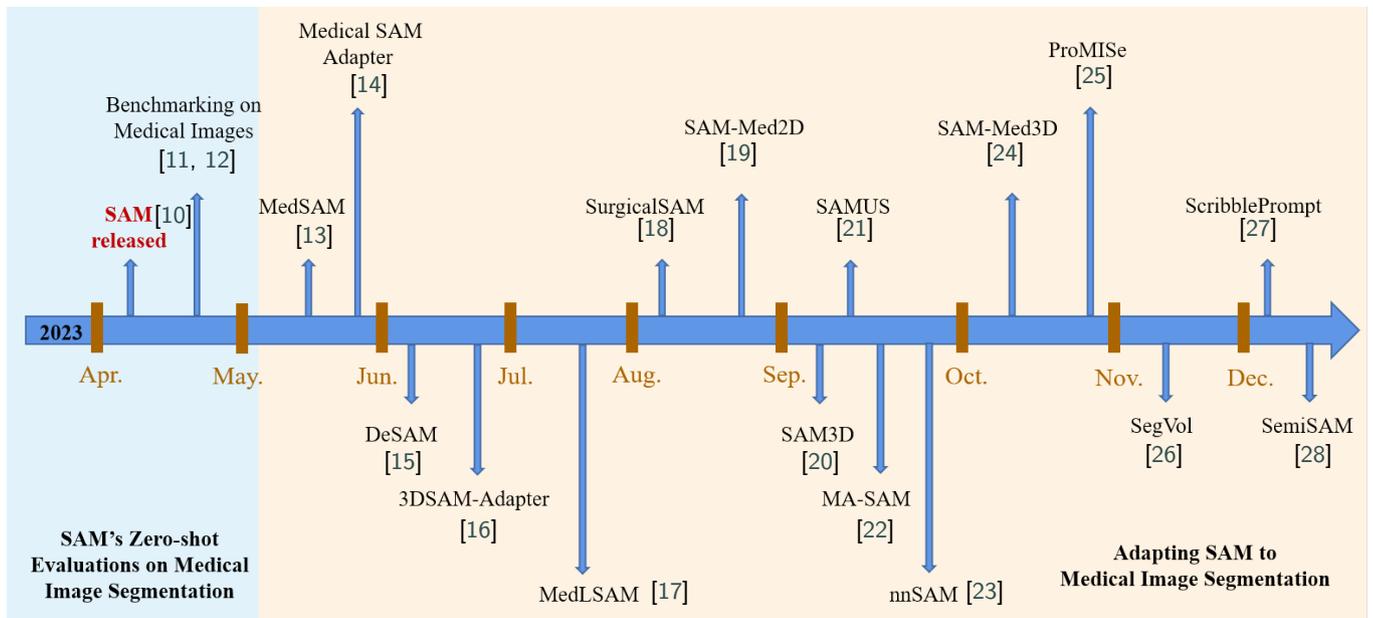

\section{Background}

\subsection{Foundation Models}

Foundation models are a rapidly growing area of artificial intelligence research aimed at developing large-scale, general-purpose language and vision models. 
These models are often trained on massive amounts of data, which allows them to learn general representations and capabilities that can be transferred to different domains and applications.
One of the most widely known foundation models is the GPT (Generative Pre-trained Transformer) series~\cite{GPT-3,GPT-4}, which demonstrated impressive capabilities and performance on a variety of natural language processing tasks such as sentence completion, question answering, and language translation.
These achievements have inspired researchers to develop large-scale foundational models to learn universal representations for computer vision tasks, which focus on capturing the cross-modal interactions between vision and language such as understanding visual concepts and details~\cite{CLIP}, generating natural language descriptions of image regions~\cite{ALIGN}, and generating images from textual descriptions~\cite{DALL-E}. 
The success of these foundation models has spawned numerous derivative works and applications spanning different industries, which have become an essential component of many AI system architectures, and their continued development promises to drive further advances in language and vision tasks~\cite{liu2023summary,awais2023foundational}.
Foundation models have also shown strong potential in solving a wide range of downstream tasks for medical image analysis and help to accelerate the development of accurate and robust models~\cite{yi2023towards,zhang2023foundation,li2023artificial}.

\subsection{Segment Anything Model}

As the first promptable foundation model for image segmentation, Segment Anything Model (SAM)~\cite{SAM-Meta} is trained on the large-scale SA-1B dataset with an unprecedented number of images and annotations, which enables the model with remarkable zero-shot generalization capability.
SAM utilizes a transformer-based architecture~\cite{attention-Nips17}, which has proven to be highly effective in natural language processing~\cite{GPT-3} and image recognition tasks~\cite{ViT2020}.
Specifically, SAM adopts an image encoder based on Vision Transformer (ViT)~\cite{ViT2020} to extract image embeddings, a prompt encoder to integrate user interactions via different prompt modes, and a lightweight mask decoder to predict segmentation masks by fusing image embeddings and prompt embeddings. 
The details of each component are presented below.

\begin{figure*}
        \centering
	\includegraphics[width=18cm]{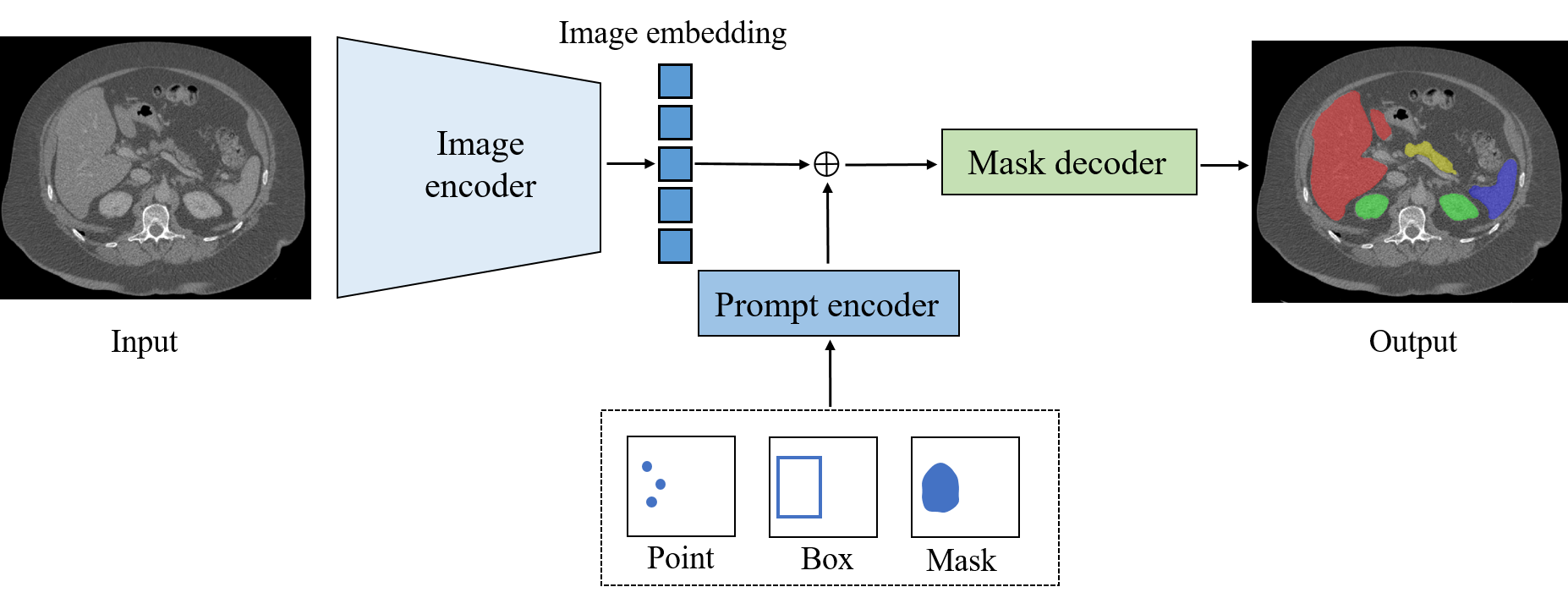}
	\caption{Overview of the architecture of Segment Anything Model (SAM), which adopts an image encoder to extract image embeddings, a prompt encoder to integrate user interactions via different prompt modes, and a mask decoder to predict segmentation masks by fusing image embeddings and prompt embeddings.}
	\label{SAM}
\end{figure*}

\textbf{Image Encoder} 
Motivated by scalability and powerful pre-training methods, SAM utilizes a ViT-based image encoder pre-trained in the scheme of Masked Autoencoder (MAE)~\cite{MAE}, which is minimally adapted to process high-resolution images.
It takes in a $1024\times1024$ image and outputs an image embedding that is $16\times$ downscaled to a 64$\times$64 feature map.

\textbf{Prompt Encoder} 
For prompt encoders, two types of prompts are considered, including sparse prompts (\ie, points, boxes) and dense prompts (\ie, masks).
SAM employs positional encoding~\cite{FourierPE-Nips20} combined with learned embeddings to represent points and boxes. 
Specifically, points are encoded by two learnable tokens for specifying foreground and background, and the bounding box is encoded by the point encoding of its top-left corner and bottom-right corner. 
Dense mask prompts, which have the same spatial resolution as the input image, are embedded using convolutions and then element-wise summation with the image embedding.

\textbf{Mask Decoder} 
The mask decoder is characterized by its lightweight design, which consists of two transformer layers with a dynamic mask prediction head and an Intersection-over-Union (IoU) score regression head. 
The mask prediction head is capable of producing three $4\times$ downscaled masks that correspond to the whole, part, and subpart of the object, respectively.

During training, the output prediction is supervised with the linear combination of focal loss~\cite{Lin2017FocalLF} and Dice loss~\cite{Milletar2016VNetFC}, and is trained for the promptable segmentation task using a mixture of geometric prompts. 
A data engine is built for label-efficient training. 
Specifically, professional annotators first label masks through interactive segmentation. 
Then, less prominent objects that are ignored in the predictions of SAM will be labeled manually. 
Finally, a fully automatic stage is conducted where confident and stable pseudo masks are selected as annotations.

\section{SAM's Zero-shot Evaluations on Medical Image Segmentation}

Despite SAM's impressive performance on natural images, it remains unclear whether it can tackle the challenge of medical image segmentation due to many intrinsic issues (\eg, structural complexity, low contrast, and inter-observer variability).
To answer this question, many studies have investigated its zero-shot performance in medical image segmentation, focusing on a diverse range of anatomical and pathological targets throughout different medical imaging modalities.
These modalities encompass both 2D medical images (\eg, X-ray, pathology, ultrasound, endoscopy, and colonoscopy) and 3D medical images (\eg, Computerized Tomography (CT), Magnetic Resonance Imaging (MRI), and Positron Emission Tomography (PET)).
Some studies particularly evaluate the effectiveness of SAM within a specific imaging modality, while others offer a comprehensive assessment across a breadth of segmentation tasks under various modalities.
In this section, we will introduce the zero-shot usage of SAM in medical image segmentation, organized according to the formats of medical imaging modalities.

\subsection{CT Image Segmentation}
A CT scan captures multiple X-ray images from different angles around the body to create a series of detailed cross-sectional slices that can visualize internal structures and abnormalities within the body, such as organs, bones, and blood vessels.
To evaluate the out-of-the-box zero-shot capabilities of SAM in segmenting abdominal organs, Roy~\etal~\cite{SAM-DKFZ-Abdomen} conduct experiments on the AMOS22 Abdominal CT Organ Segmentation dataset~\cite{ji2022amos}.
They produce different settings of randomly selected points and jittered bounding boxes from the segmentation masks to simulate various degrees of user inaccuracy. 
Their results show that SAM with point prompts underperforms state-of-the-art (SOTA) performance with an average Dice Similarity Coefficient (DSC) decrease ranging from 20.3\% to 40.9\% while using box prompts can obtain highly competitive performance even with moderate jittering.
To assess SAM's performance in segmenting tumors, Hu~\etal~\cite{SAM-LiverTumor} conduct experiments on multi-phase liver tumor segmentation from contrast-enhanced computed tomography (CECT) volumes. 
Their results demonstrate that the more prompt points used for segmentation, the better performance SAM can achieve.
However, there is still a large performance gap with classic U-Net architectures~\cite{U-Net}.

\subsection{MRI Image Segmentation} MRI is a non-invasive imaging technique that utilizes powerful magnets and radio waves to generate high-resolution cross-sectional views of internal anatomical structures, including brains, joints, and other soft tissues.
Since MRI plays a significant role in brain visualization, Mohapatra~\etal~\cite{SAM-BrainMR} compared SAM with Brain Extraction Tool (BET), which is a widely used and current gold standard technique for brain extraction and segmentation. 
They conduct experiments on a variety of brain scans with varying image qualities, MR sequences, and brain lesions.
Their results show that SAM can obtain comparable or even better performance, demonstrating its potential to emerge as an efficient tool for brain extraction and segmentation. 
Zhang~\etal~\cite{SAM-BraTS} also evaluate the performance of SAM on brain tumor segmentation and validate that there is still a gap between SAM and current SOTA methods if model fine-tuning is not implemented.

\begin{figure*}
	\includegraphics[width=18cm]{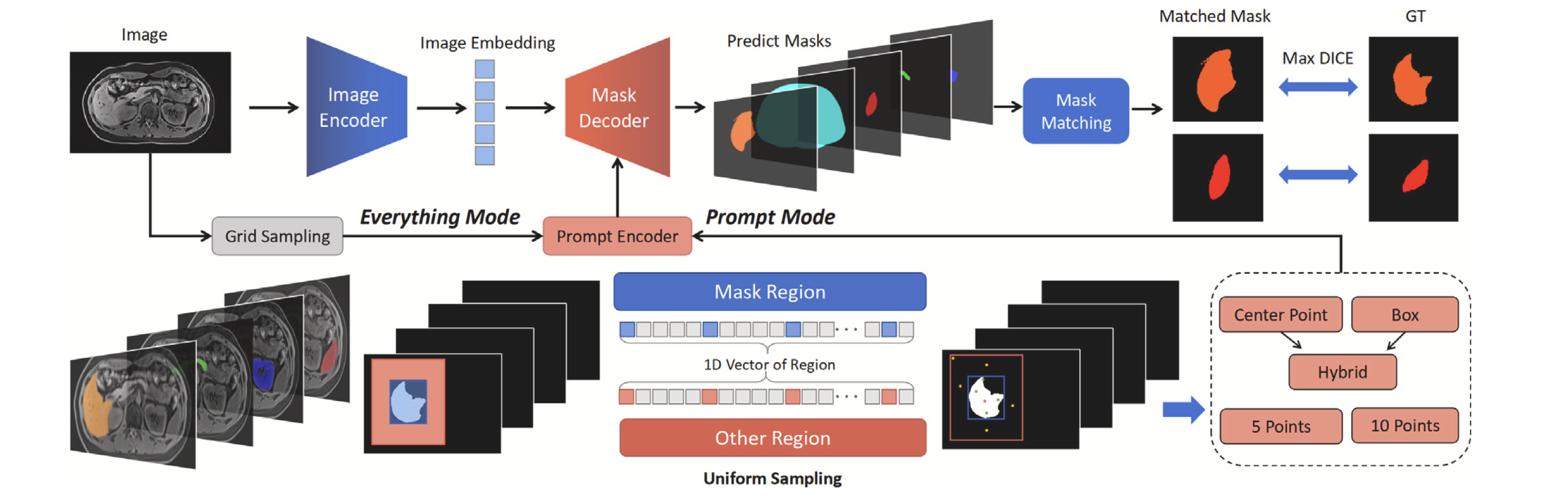}
	\caption{SAM's zero-shot evaluation pipeline on medical image segmentation in a large-scale empirical study~\cite{SAM-SZU}.}
	\label{SAM-eval}
\end{figure*}

\subsection{Pathological Image Segmentation} 
Pathological images are typically captured using microscopy techniques to depict abnormal tissue structures, cellular changes, or pathological conditions within the human body, especially in oncology. 
Deng~\etal~\cite{SAM-pathology} evaluate SAM on tumor segmentation, non-tumor tissue segmentation, and cell nuclei segmentation on whole slide imaging (WSI) data. 
By conducting several prompt settings including a single positive point, 20 points (10 positives and 10 negatives), and total points/boxes, the results suggest that SAM can achieve remarkable segmentation performance for large connected objects but fails for dense instance object segmentation, even with 20 prompts (clicks/boxes) on each image. 
Possible reasons encompass the notably larger image resolution of WSI data \textit{vs.} the training image resolution of SAM, as well as multiple scales of different tissue types for digital pathology.

\subsection{Colonoscopic Image Segmentation} 
A colonoscopic image is obtained during a colonoscopy procedure by inserting a colonoscope into the rectum to examine the inner lining of the colon and rectum.
The colonoscopic images can provide a detailed view of the colon's lining, allowing clinicians to identify abnormalities like polyps and inflammation.
Zhou~\etal~\cite{SAM-Polyps} evaluate the performance of SAM with unprompted setting in segmenting polyps from colonoscopy images on five benchmark datasets~\cite{silva2014toward,bernal2015wm,tajbakhsh2015automated,vazquez2017benchmark,jha2020kvasir}. 
The experimental findings indicate a noticeably lower performance of SAM compared to SOTA methods with an average Dice Similarity Coefficient (DSC) drop ranging from 14.4\% to 36.9\%.
SAM fails to achieve satisfactory performance when directly applying polyp segmentation task due to blurred boundaries between a polyp and its surrounding mucosa, suggesting the necessity for the involvement of prompts or an adaptation method to improve performance.

\subsection{Endoscopic Image Segmentation}
An endoscopic image refers to a visual representation captured through endoscopy, which is a minimally invasive medical procedure for examining the interior of a hollow organ or cavity within the body. 
Endoscopy is typically used in robot-assisted surgery where the segmentation of surgical instruments is essential for instrument tracking and position estimation.
Wang~\etal~\cite{SAM-RS} conduct an assessment of SAM's performance on two public datasets of endoscopic surgical instrument segmentation, \ie, EndoVis17~\cite{allan20192017} and EndoVis18~\cite{allan20202018}. 
The experimental results demonstrate shortcomings in SAM's ability to accurately segment entire instruments when utilizing both point-based prompts and unprompted settings.
Notably, SAM exhibits deficiencies in predicting certain instrument parts, especially for overlapping instruments when prompts are weak. 
Additionally, SAM faces challenges in identifying instruments within complex surgical scenarios characterized by blood, reflection, blur, and shade.

\subsection{Segmentation in Multiple Modalities} 
Instead of evaluating SAM on a single imaging modality, He~\etal~\cite{SAM-Meds} conduct a large-scale empirical study to evaluate the accuracy of SAM on 12 public medical image segmentation datasets covering different organs including the brain, breast, chest, lung, skin, liver, bowel, pancreas, and prostate. 
Their evaluation spans various imaging modalities (\eg, 2D X-ray, 2D ultrasound, 3D MRI, and 3D CT), and encompasses different health conditions including both normal and abnormal cases. 
The results illustrate that SAM currently falls short of accuracy when directly applied to medical images without fine-tuning, and its performance can be influenced by multiple factors such as dimensionality, modality, size, and contrast.
Similarly, Mazurowski~\etal~\cite{SAM-Empirical} conduct an extensive evaluation of SAM on a collection of 11 medical image segmentation datasets of different modalities by generating point prompts to simulate interactive segmentation. 
The observed performance indicates high accuracy for well-circumscribed objects with unambiguous prompts but lower precision in segmenting tumors.
Cheng~\etal~\cite{SAM-MI} evaluate the performance of SAM on 12 public medical image datasets representing various organs and modalities under three prompt modes, including auto-prompt, box-prompt, and point-prompt modes. 
The experimental results reveal variations in SAM's performance across different datasets, with the box-prompt mode without jitters proving to be the most effective for utilizing SAM in zero-shot medical image segmentation.
Zhang~\etal~\cite{SAM-RO} evaluate the performance of SAM in clinical radiotherapy by segmenting prostate, lung, gastrointestinal, and head\&neck, which are major treatment sites in radiation oncology. 
The evaluation underscored SAM's capability to delineate large, distinct organs but highlighted challenges in segmenting smaller, intricate structures, particularly when faced with ambiguous prompts.
To fully validate SAM’s performance in medical
data, Huang~\etal~\cite{SAM-SZU} curated and organized 52 open-source datasets to create COSMOS 1050K, a large-scale medical segmentation dataset featuring 18 modalities, 84 objects, 125 object-modality paired targets, 1050K 2D images, and 6033K masks.
Comprehensive experiments are conducted on different SAM prompt modes including everything modes, point-based and box-based prompt modes.
The experimental results validate that SAM exhibits higher performance with manual prompts, \ie, points and boxes, for object perception in medical images compared to everything mode.
Additionally, it is observed that the addition of negative points, which should theoretically improve performance, will slightly decrease performance in some tasks, especially when background objects resemble foreground ones.
This discovery underscores the importance of judiciously selecting point prompts based on domain knowledge to achieve stable performance improvement.

\begin{figure*}
	\includegraphics[width=18cm]{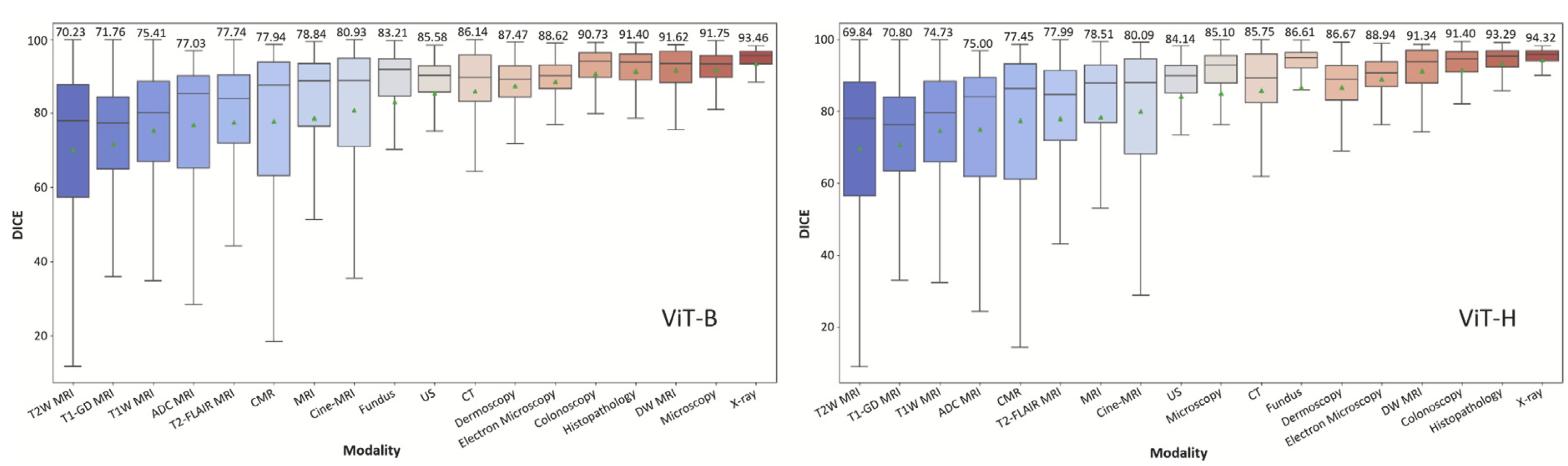}
	\caption{Quatitative segmentation results of SAM on 18 different imaging modalities in a large-scale empirical study~\cite{SAM-SZU}.}
	\label{Comparison}
\end{figure*}

\subsection{Summary} 
In this section, we review recent studies to explore SAM's potential for zero-shot transfer to different medical image segmentation tasks, comparing its performance with existing domain-specific segmentation methods.
Generally, SAM necessitates substantial human interaction to achieve moderate segmentation performance, which requires only a few points or bounding box prompts. 
The evaluation results across various datasets indicate SAM's limited generalization ability when directly applied to medical image segmentation, which varies significantly across different datasets and tasks.
While SAM demonstrates remarkable performance comparable to SOTA methods in discerning well-circumscribed objects in certain imaging modalities, it exhibits imperfections or total failures in more challenging situations.
This is particularly evident when dealing with segmentation targets featuring weak boundaries, low contrast, small size, and irregular shapes, aligning with findings from other investigations~\cite{SAM-ConcealedScenes,SAM-Realworld}.
For most medical image segmentation scenarios, SAM's subpar segmentation performance falls short of the requirements for further applications, particularly in some tasks that demand extremely high accuracy. 
The SA-1B dataset, the training data of SAM, is primarily composed of natural images with strong edge information and poses a significant dissimilarity from medical images. 
Consequently, directly applying SAM without fine-tuning or re-training to previously unseen and challenging medical image segmentation may yield limited performance.

\section{Adapting SAM to Medical Image Segmentation}
Given the persisting challenges in SAM's zero-shot transfer to medical image segmentation, an alternative research direction has emerged, emphasizing the improved adaptation of SAM to diverse medical image segmentation tasks.
Notably, considerable attention has been devoted to refining SAM for both 2D and 3D imaging modalities, which includes fine-tuning different SAM modules and training architectures analogous to SAM from scratch. 
These endeavors aim to boost SAM's performance in medical image segmentation tasks, enabling it to better accommodate diverse data characteristics and complexities. 

\begin{figure}
	\includegraphics[width=8cm]{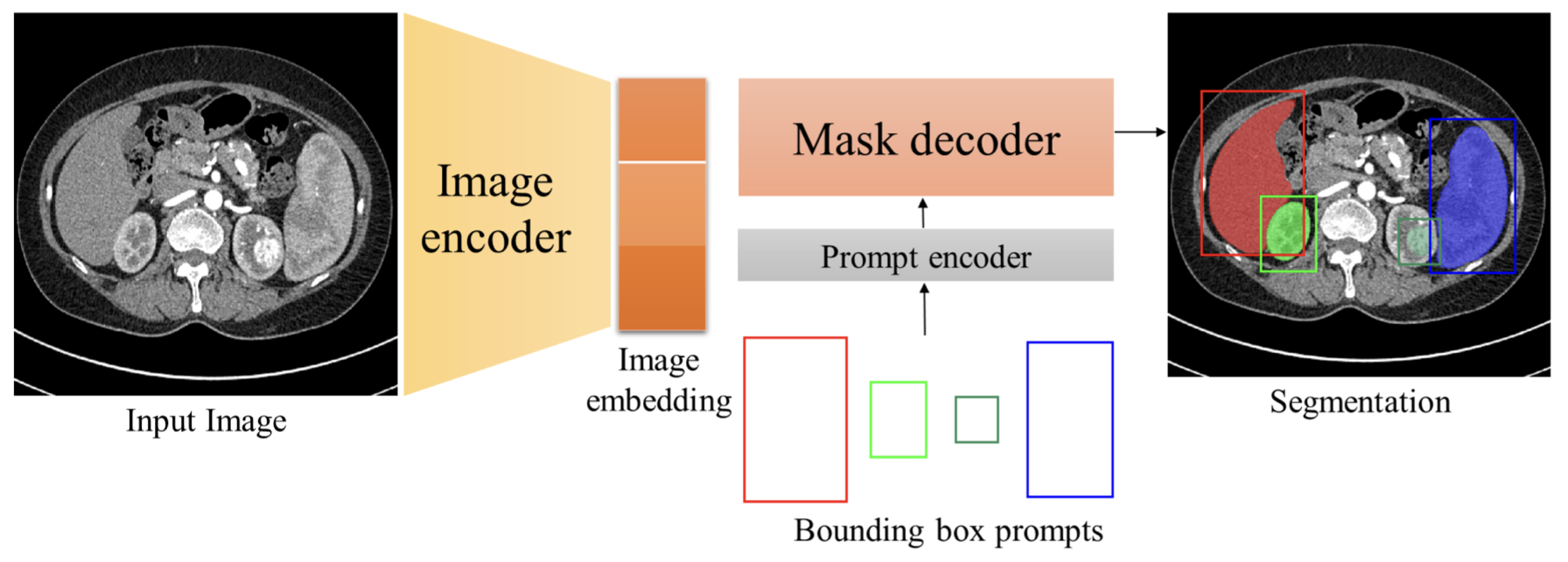}
	\caption{MedSAM~\cite{MedSAM} adapts SAM for medical image segmentation by freezing the prompt encoder while fine-tuning the image encoder and the mask decoder.}
	\label{MedSAM}
\end{figure}

\begin{figure}
	\includegraphics[width=8cm]{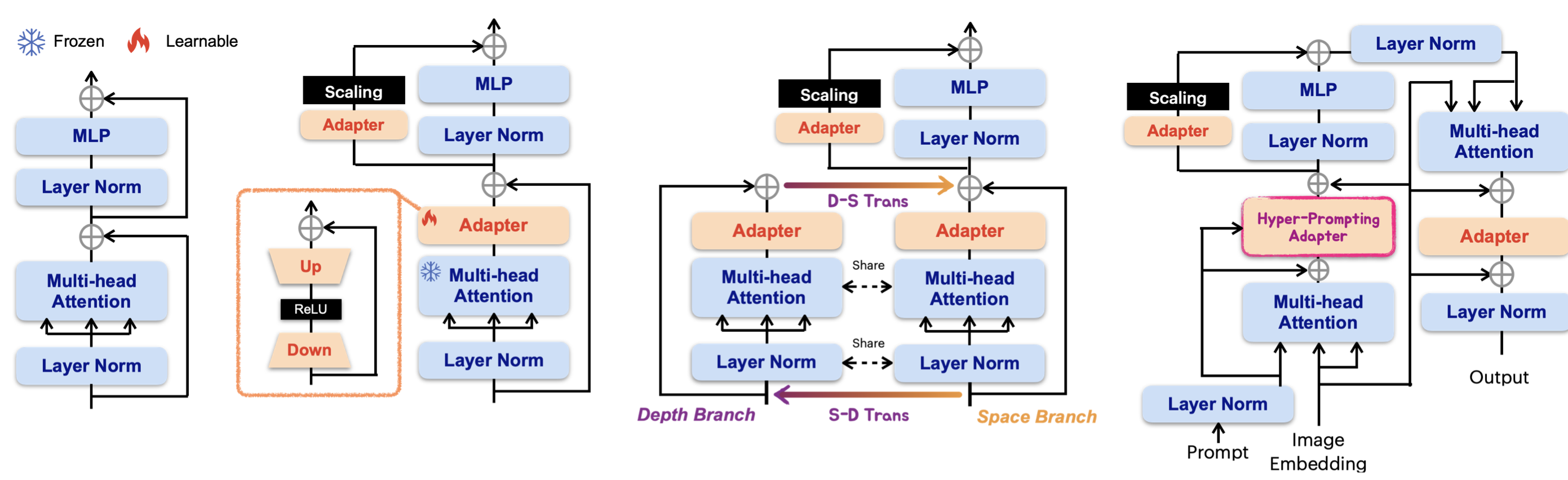}
	\caption{Medical SAM Adapter (Med-SA)~\cite{Med-SAM-Adapter} adapts SAM for medical image segmentation in a parameter-efficient way.}
	\label{MSA}
\end{figure}

\subsection{Fine-tuning on Medical Images}

To improve the unsatisfactory performance of SAM on medical image segmentation tasks, a direct and intuitive approach is to fine-tune SAM on medical images, including full fine-tuning and parameter-efficient fine-tuning.

\subsubsection{Full Fine-tuning} 
The most straightforward approach to adapting SAM for medical image segmentation is to directly fine-tune SAM on the specific task at hand.
Hu~\etal~\cite{skinSAM} present a fine-tuning validation of SAM for skin cancer segmentation, demonstrating a substantial improvement in the DSC score from 81.25\% to 88.79\%.
Li~\etal~\cite{PolypSAM} propose PolypSAM by fine-tuning all
components of SAM for polyp segmentation, which achieves outstanding performance on five public datasets with the DSC scores all above 88\%. 
MedSAM~\cite{MedSAM} is introduced for universal medical image segmentation, which adapts from SAM on an unprecedented scale by curating a diverse and comprehensive dataset containing more than one million medical image-mask pairs of 11 modalities.
MedSAM achieved median DSC scores of 94.0\%, 94.4\%, 81.5\%, and 98.4\% for the segmentation tasks involving intracranial hemorrhage CT, glioma MR T1, pneumothorax CXR, and polyp endoscopy images, respectively, surpassing the performance of the U-Net specialist models.
However, MedSAM faces challenges in segmenting vessel-like branching structures due to the potential ambiguity in bounding box prompts in such scenarios.
Moreover, it only processes 3D images as a series of 2D slices instead of the volumes.

\subsubsection{Parameter-efficient Fine-tuning} 

\begin{figure}
	\includegraphics[width=8cm]{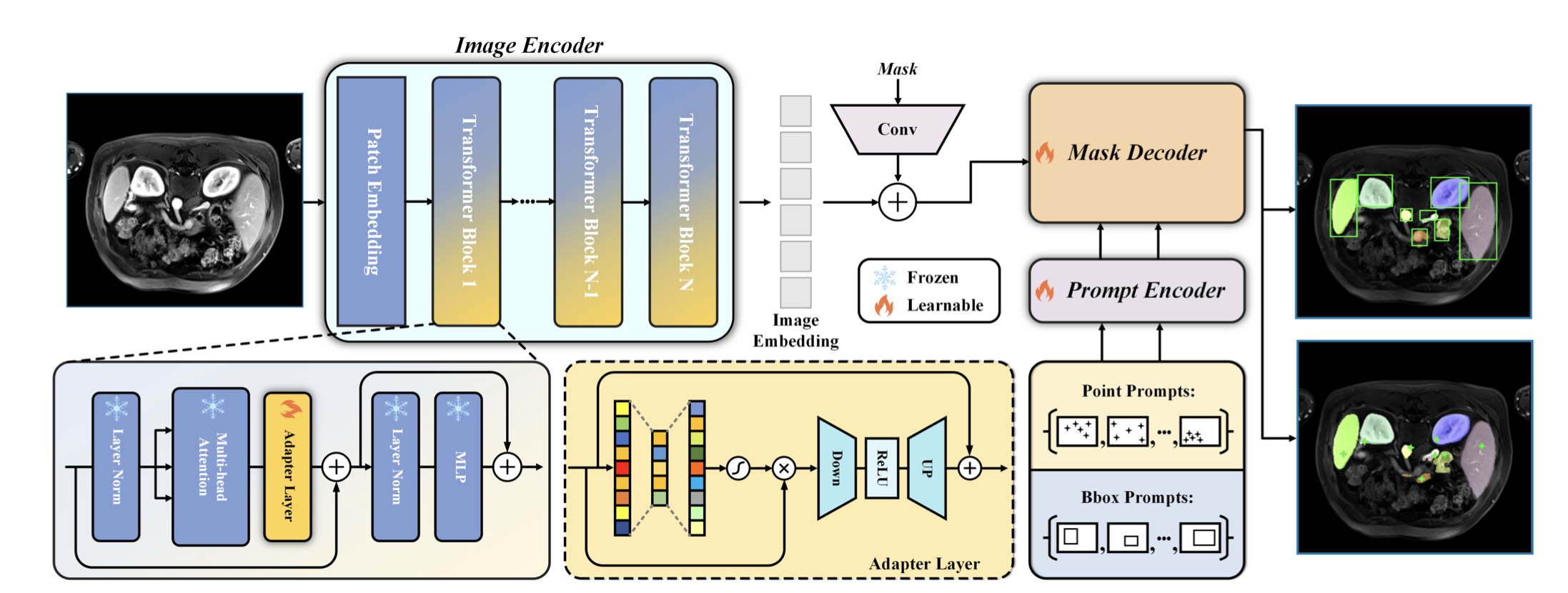}
	\caption{SAM-Med2D~\cite{SAM-Med2D} freezes the image encoder while introducing learnable adapter layers, fine-tuning the prompt encoder, and updating the mask decoder during training.}
	\label{SAM-Med2D}
\end{figure}

Updating all parameters of SAM is a time-consuming, computationally intensive, and challenging process, making it less feasible for widespread deployment.
Consequently, many researchers focus on fine-tuning a small fraction of the parameters of SAM using various parameter-efficient fine-tuning (PEFT) techniques.
Rather than fully adjusting all parameters, Wu~\etal~\cite{Med-SAM-Adapter} propose Medical SAM Adapter (Med-SA), which maintains the pre-trained SAM parameters frozen while integrating low-rank adaptation (LoRA) modules~\cite{Lora} to the designated positions. 
The extensive experiments on 17 medical image segmentation tasks across 5 different modalities showcase Med-SA’s superiority over SAM and previous SOTA methods.
Similarly, SAMed~\cite{SAMed} applies LoRA modules to the pre-trained SAM image encoder and fine-tunes it together with the prompt encoder and the mask decoder on Synapse multi-organ segmentation dataset. 
SAMed only updates a small fraction of the SAM parameters, but it can achieve a DSC score of 81.88\% on par with the SOTA methods.
Feng~\etal\cite{feng2023cheap} introduce an efficient and practical approach for fine-tuning SAM using a limited number of exemplars, which combines an exemplar-guided synthesis module and LoRA fine-tuning strategy, demonstrating SAM's effective alignment within the medical domain even with few labeled data.
Paranjape~\etal~\cite{paranjape2023adaptivesam} propose AdaptiveSAM, an adaptive modification to efficiently adapt SAM to new datasets and enable text-prompted segmentation in the medical domain. 
It employs bias-tuning with a significantly smaller number of trainable parameters than SAM while utilizing free-form text prompts for object segmentation.
The experiments show that AdaptiveSAM outperforms current SOTA methods on various medical imaging datasets including surgery, ultrasound, and X-ray.
To bridge the large domain gap between natural images and medical images, Cheng~\etal~\cite{SAM-Med2D} introduce SAM-Med2D, the most comprehensive study on applying SAM to medical 2D images by incorporating learnable adapter layers in the image encoder, fine-tuning the prompt encoder, and updating the mask decoder through interactive training.
They collected and curated a medical image segmentation dataset comprising over 4.6M images and 19.7M masks.
The comprehensive evaluation and analysis are conducted to investigate SAM-Med2D's performance across various modalities, anatomical structures, and organs, as well as the generalization capability on 9 MICCAI 2023 challenge datasets, demonstrating its significantly superior performance and generalization capability compared to SAM.

\subsection{Auto-prompting Adaptation}
While effective in certain scenarios, existing SAM adaptions require relative high-quality typical prompts of SAM (\ie, points, boxes, and masks) to achieve acceptable performance in medical image segmentation tasks.
In most of these efforts, prompts are generated from the ground truth during testing~\cite{MedSAM,feng2023cheap,Med-SAM-Adapter}. 
However, creating accurate and reliable prompts still requires domain-specific knowledge from medical experts, which may not be available. This is particularly challenging in the context of universal medical image segmentation involving numerous classes. Additionally, low-quality prompts induced by noisy annotation can significantly compromise the accuracy of segmentation.
Consequently, the quest for an auto-prompting mechanism aims to establish a robust and adaptive framework that mitigates the variability in SAM's performance and contributes to more reliable and accurate outcomes in medical image segmentation.

\subsubsection{Prompts Auto-generation}

To enable auto-prompting, a straightforward approach is to utilize a localization framework to generate input prompts for SAM.
To segment regions of interest (ROI) in diverse medical
imaging datasets, Pandey~\etal~\cite{pandey2023comprehensive} employ a YOLOv8 model to obtain ROI bounding boxes as SAM's input prompt for fully automatic medical image segmentation.
MedLSAM~\cite{MedLSAM} applies a few-shot localization process by identifying 3D bounding boxes that enclose any anatomical structure of interest in 3D medical images based on the assumption that images with locally similar pixel distributions correspond to the same region in different individuals.
Subsequently, 2D boxes are derived from the projection of the 3D box onto each slice, guiding SAM to automatically segment the target anatomy.
Anand~\etal~\cite{anand2023one} propose a one-shot localization and segmentation framework to leverage the correspondence with respect to a template image to prompt SAM.
They utilize the pre-trained ViT-based foundation models to extract dense features from the template image,

\begin{figure}
	\includegraphics[width=8cm]{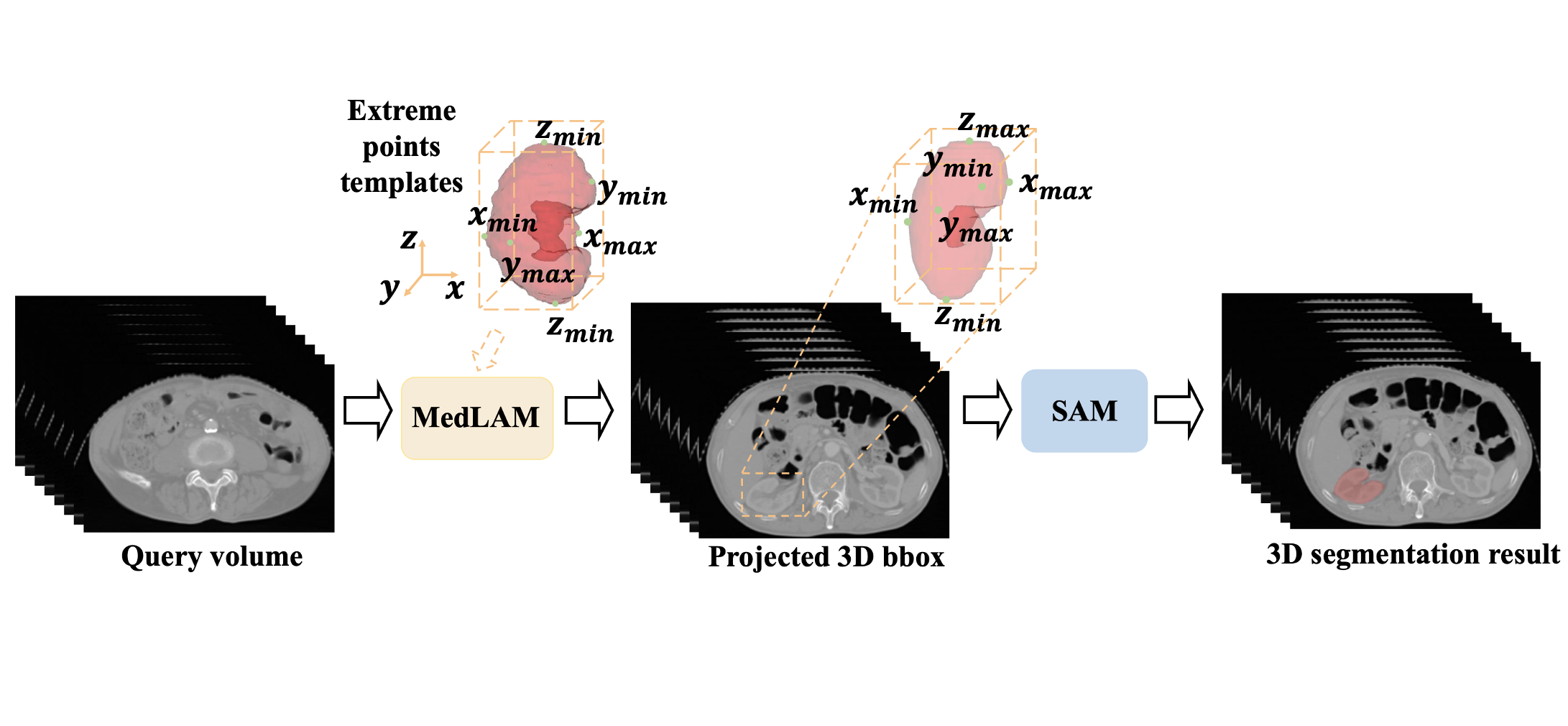}
	\caption{Pipeline of MedLSAM~\cite{MedLSAM} for auto-generation of box prompts in 3D medical image segmentation.}
	\label{MedLSAM}
\end{figure}

\begin{figure}
	\includegraphics[width=8cm]{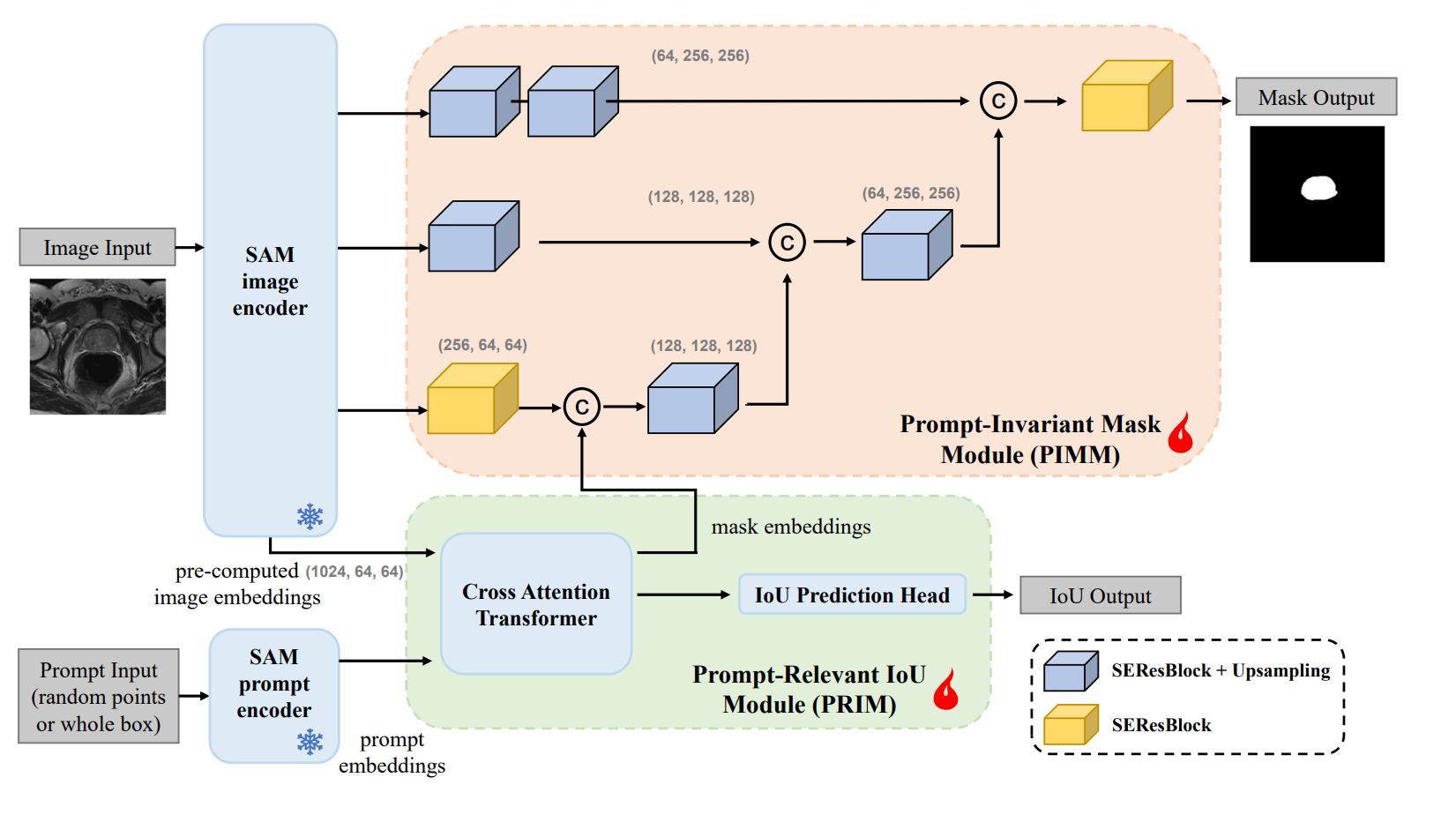}
	\caption{Decoupling Segment Anything Model (DeSAM)~\cite{DeSAM} decompose the mask decoder into a Prompt-Relevant IoU Module (PRIM) to generate mask embeddings and a Prompt-Invariant Mask Module (PIMM) to fuse the image embeddings with the generated mask embeddings.}
	\label{DeSAM}
\end{figure}

\subsubsection{Learnable Prompts}
AutoSAM~\etal~\cite{AutoSAM} involves the training of an auxiliary prompt encoder to generate a surrogate prompt without further fine-tuning SAM. 
The auxiliary prompt encoder extracts the features of the input image itself as the conditional prompts, expanding beyond the typical prompts. 
Through this strategy, SAM is turned into a fully auto-prompting manner, eliminating the necessity for manual prompts.
AutoSAM demonstrates SOTA results across various medical benchmarks, showcasing its superior performance in medical image segmentation tasks.
The all-in-SAM pipeline~\cite{All-in-SAM} first utilizes the pre-trained SAM to generate pixel-level annotations from weak prompts and then uses them to finetune SAM following the strategy from~\cite{chen2023sam}.
Such a pipeline does not require manual prompts during the inference stage, surpassing previous SOTA methods in nuclei segmentation and achieving competitive performance compared to using strong pixel-wise annotated data.
Gao~\etal~\cite{DeSAM} propose Decoupling Segment Anything Model (DeSAM) to address the coupling effect of poor prompts and mask segmentation in medical image segmentation.
They decouple SAM’s mask decoder to carry out two sub-tasks, including a Prompt-Relevant IoU Module (PRIM) to generate mask embeddings based on the given prompt, and a Prompt-Invariant Mask Module (PIMM) to fuse image embeddings with the mask embeddings for the final segmentation mask.
The extensive experiments indicate that DeSAM can improve the robustness of SAM's fully automatic mode by an average DSC score of 8.96\% in dealing with domain shifts across different clinical sites.
Yue~\etal~\cite{SurgicalSAM} introduces SurgicalSAM, which integrates surgical-specific information with SAM's pre-trained knowledge for improved generalization by utilizing a lightweight prototype-based class prompt encoder for fine-tuning and contrastive prototype learning for more accurate class prompting.
The results of extensive experiments on two public datasets demonstrate that SurgicalSAM achieves SOTA performance while only requiring a small number of tunable parameters.

\subsubsection{Enhancing Reliability Against Prompts with Uncertainty}
Given the sensitivity of SAM to the input prompt, the estimation of uncertainty becomes essential to guarantee the reliability of segmentation results.
This is particularly critical in medical imaging where segmentation accuracy plays a significant role in clinical procedures.
Xu~\etal~\cite{xu2023eviprompt} propose a training-free prompt generation method based on uncertainty estimation named EviPrompt to automatically generate prompts for SAM in segmenting medical images without the interaction of clinical experts, which requires only a single medical image-annotation pair as a reference.
Deng~\etal~\cite{SAM-U} propose a multi-box prompt-triggered uncertainty estimation as a test-time augmentation technique for SAM in segmenting fundus images.
They generate different predictions from multiple box prompts, estimate the distribution via Monte Carlo simulation, and establish an uncertainty map that offers guidance for potential segmentation error, thus enhancing SAM's robustness against different prompts. 
Zhang~\etal~\cite{UR-SAM} propose UR-SAM, an uncertainty rectified SAM framework to enhance the robustness and reliability for auto-prompting medical image segmentation by estimating the uncertainty maps and utilizing the uncertainty to rectify possible error and improve the segmentation result.
Their experiments on two public 3D medical datasets covering the segmentation of 35 organs demonstrate that estimating and utilizing uncertainty can improve the segmentation performance with up to 10.7\% and 13.8\% in DSC scores without manual prompts.
Consequently, the integration of uncertainty can enhance SAM's robustness against various prompts.
Estimated uncertainty not only aids in identifying potential segmentation errors but also provides valuable guidance for clinicians, thereby enhancing the overall segmentation reliability and facilitating further applications.

\begin{figure}
	\includegraphics[width=8cm]{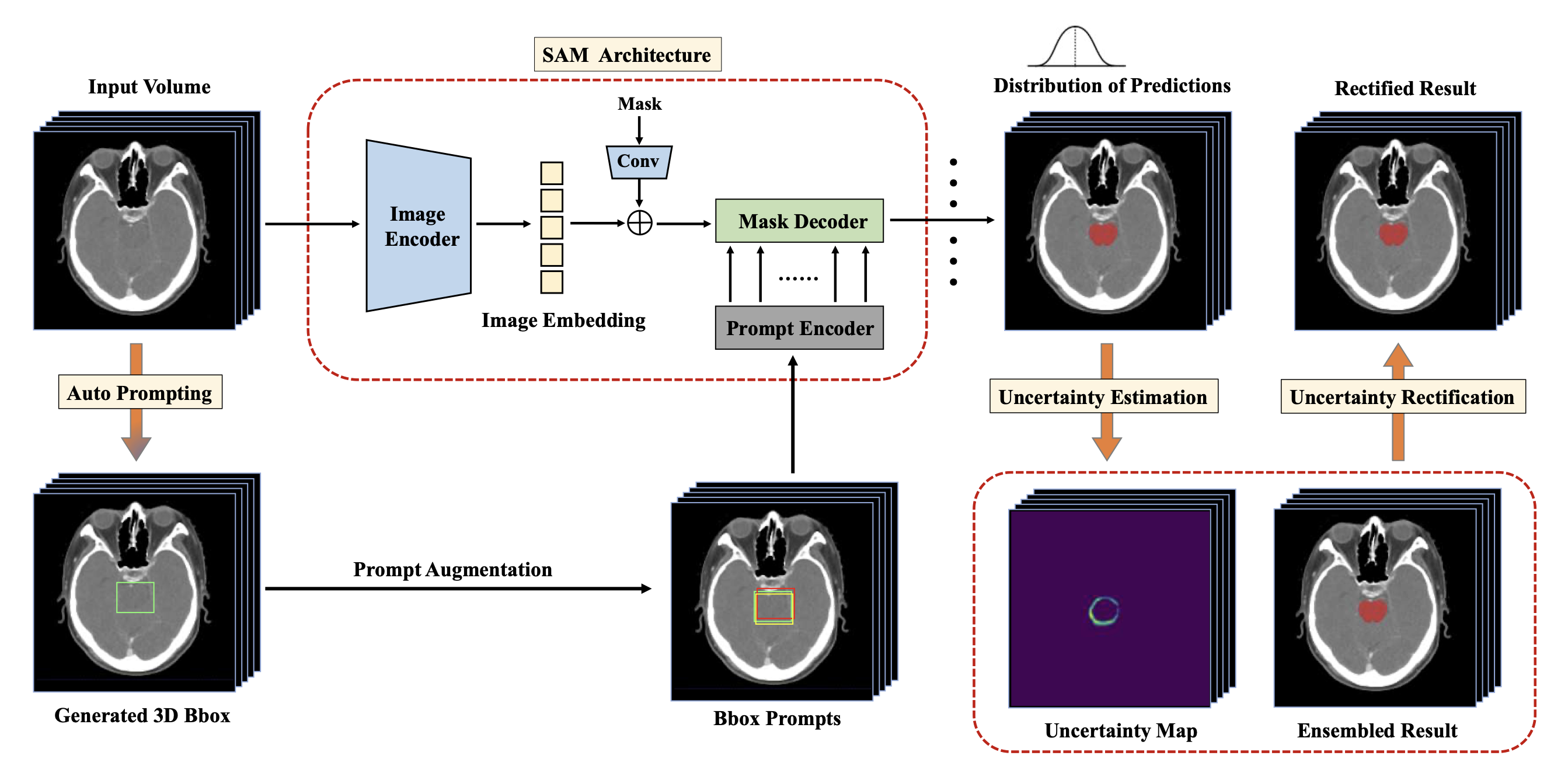}
	\caption{UR-SAM~\cite{UR-SAM} enhances the reliability of auto-prompting medical image segmentation by estimating and utilizing uncertainty for rectification of segmentation results.}
	\label{UR-SAM}
\end{figure}

\subsection{Framework Modification}
Since SAM is a powerful foundational model pre-trained on natural image segmentation, many endeavors have sought to harness the capabilities of SAM by either modifying its existing framework or seamlessly integrating it into novel training schemes towards the construction of advanced medical image segmentation models. 

\subsubsection{Synergy in Training Segmentation Models}
Zhang~\etal~\cite{SAM-Path} propose a fine-tuning framework SAM-Path to adapt SAM for semantic segmentation in digital pathology.
SAM-Path introduces trainable class prompts for the targets of interest and a pre-trained pathology encoder for incorporating domain-specific knowledge to compensate for the lack of comprehensive pathology datasets used in SAM training. 
The experiments on the CRAG dataset show a relative improvement of 27.52\% in the DSC score compared to the vanilla SAM with manual prompts.
Chai~\etal~\cite{chai2023ladder} utilize a ladder fine-tuning scheme that combines a complementary CNN encoder with the standard SAM architecture and only focuses on fine-tuning the additional CNN and SAM decoder for computational resource and training time reduction. 
Li~\etal~\cite{li2023nnsam} present nnSAM, which synergistically integrates the pre-trained SAM model as a plug-and-play module with
nnU-Net to achieve more accurate and robust medical image segmentation.
Zhang~\etal~\cite{IA-SAM} propose SAMAug which directly utilizes the segmentation masks generated by SAM to augment the raw inputs of commonly-used medical image segmentation models (\eg, U-Net).
The experiments on two datasets show that although SAM may not generate high-quality segmentation for medical images, its generated masks and features are still useful for training better medical image segmentation models. 
Lin~\etal~\cite{lin2023samus} propose SAMUS by introducing a parallel CNN branch to inject local features into the ViT encoder, a position adapter and a feature adapter to adapt SAM from large-size inputs to small-size inputs for more clinical-friendly deployment. 
A comprehensive ultrasound dataset comprising 30k images and 69k masks with six object categories is collected and curated for evaluation, demonstrating its superiority against the SOTA task-specific models and universal foundation models in ultrasound image segmentation.

\begin{figure}
	\includegraphics[width=8cm]{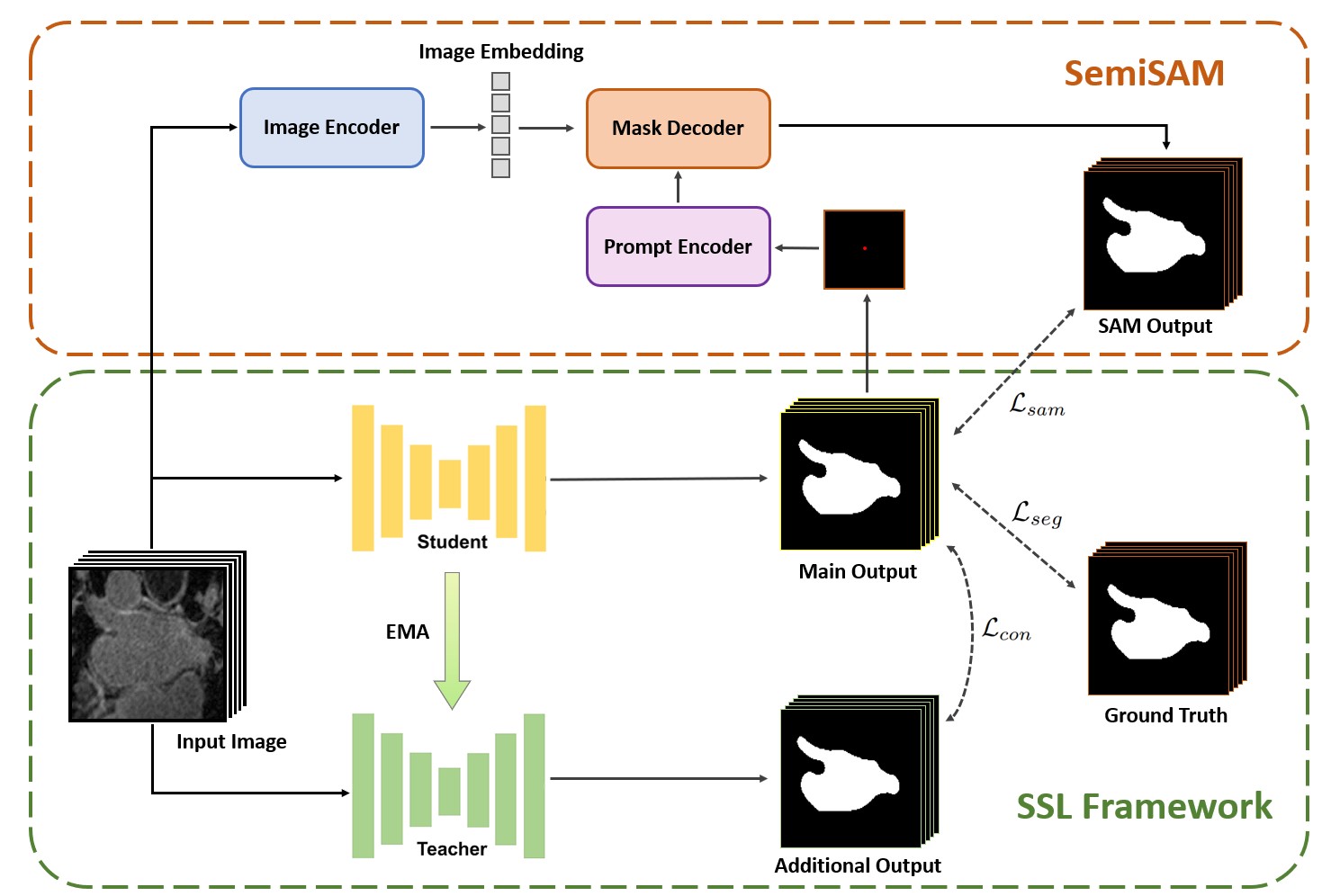}
	\caption{SemiSAM~\cite{zhang2023semisam} explores the usage of SAM as an additional supervision branch to assist in the learning procedure of a semi-supervised framework.}
	\label{SemiSAM}
\end{figure}

\subsubsection{Facilitating Annotation-efficient Learning}
Due to the high annotation cost of medical image segmentation as it necessitates the expertise of experienced clinical professionals, significant efforts have been devoted to annotation-efficient learning, like semi-supervised learning and weakly supervised learning.
As a reliable pseudo-label generator, SAM openes up new opportunities to guide the segmentation task when manually annotated images are scarce.
Zhang~\etal~\cite{zhang2023samdsk} propose an iterative semi-supervised method that combines SAM-generated segmentation proposals with pixel-level and image-level domain-specific knowledge for repeatedly constructing annotations of unlabeled images.
To produce reliable pseudo labels, Li~\etal~\cite{li2023segment} leverage the pre-trained SAM to conduct predictions consistent with the generated pseudo labels and select the reliable pseudo labels to further boost the existing semi-supervised segmentation model, which achieves 6.84\% and 10.76\% improvement over the advanced two baselines respectively on 5\% of labeled data from the publicly available ACDC dataset.
Instead of generating pseudo labels, Zhang~\etal~\cite{zhang2023semisam} propose a semi-supervised framework named SemiSAM, where the segmentation model trained with domain knowledge provides localization information (\ie, input prompts) to SAM while SAM serves as an additional supervision branch to assist in the consistency learning.
The experimental results on left atrium MRI segmentation dataset demonstrate that SemiSAM achieves significant improvements especially when labeled data are extremely limited.
\cite{li2023leverage} propose to minimize the labeling efforts by utilizing weak box annotations instead of pixel-level delineation in the context of high-resolution whole slide imaging.

\subsection{Towards 3D Medical Images}
Directly leveraging the pre-trained SAM with inherent 2D architecture often leads to sub-optimal results in 3D medical image segmentation because slice-wise (2D) segmentation via transfer learning usually discards important depth-related spatial context in 3D medical images, which is extremely important for identification of some objects so as to ensure accurate segmentation~\cite{2-5D}.
To address this issue, many studies have undertaken specific modifications and enhancements to enable SAM to effectively process 3D medical image modalities.

\begin{figure}
	\includegraphics[width=8cm]{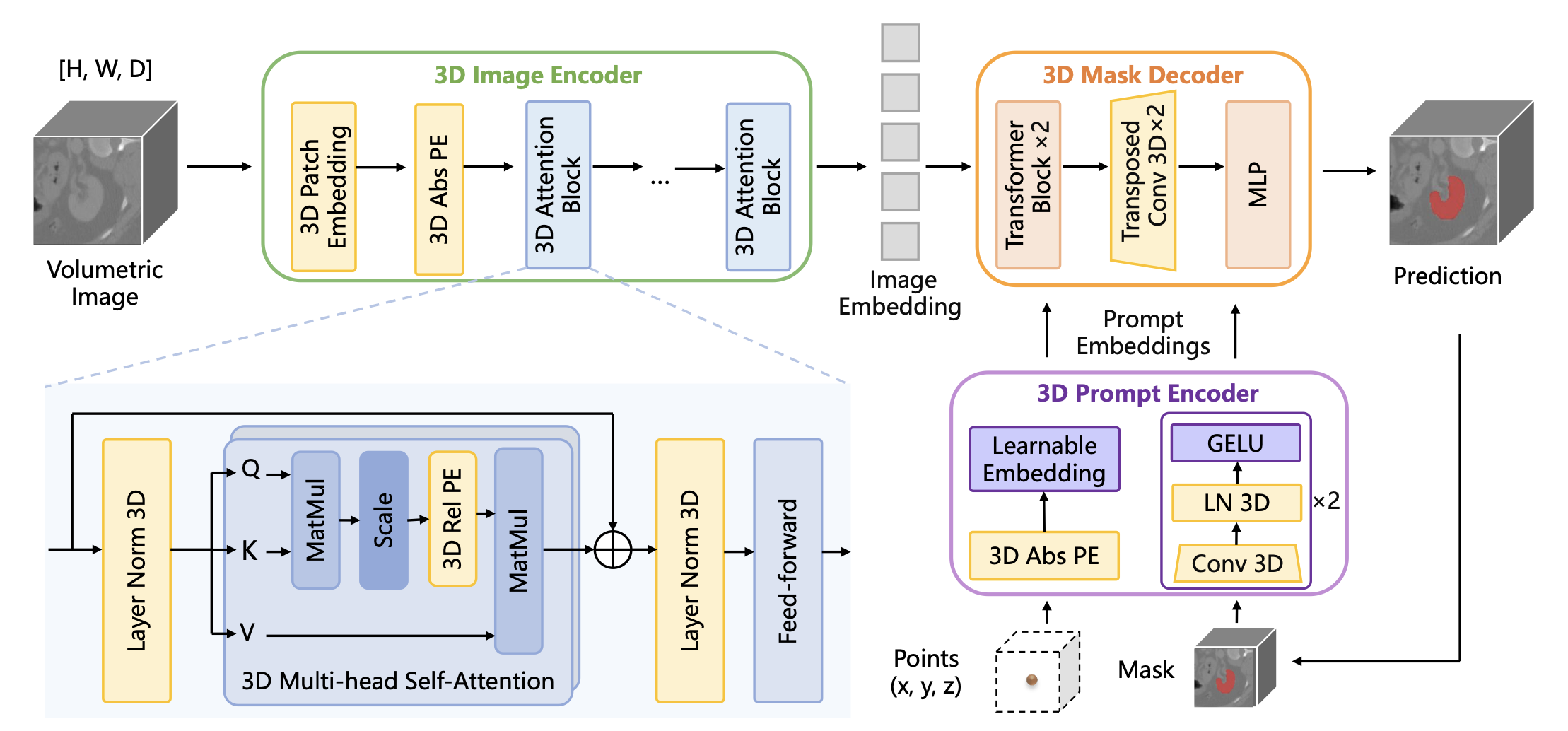}
	\caption{SAM-Med3D~\cite{SAM-Med3D} transforms original 2D components of SAM into their 3D counterparts, encompassing a 3D image encoder, 3D prompt encoder, and 3D mask decoder.}
	\label{SAM-Med3D}
\end{figure}

\subsubsection{Adaptation from 2D to 3D}
To achieve 2D to 3D adaptation, Medical SAM Adapter (Med-SA)~\cite{Med-SAM-Adapter} introduce the Space-Depth Transpose (SD-Trans) technique where a bifurcated attention mechanism is utilized by capturing spatial correlations in one branch and depth correlations in another. 
Gong~\etal~\cite{3DSAM-adapter} propose 3DSAM-adapter, a carefully designed modification of SAM architecture to support volumetric medical image segmentation with only 16.96\% tunable parameters (including newly added parameters) of the original model.
The experiment results show that 3DSAM-adapter significantly outperforms nnU-Net~\cite{isensee2021nnu} on three datasets (by 8.25\% for kidney tumor, 29.87\% for pancreas tumor, and 10.11\% for colon cancer).
Chen~\etal~\cite{MA-SAM} introduce a modality-agnostic SAM adaptation framework (MA-SAM) applicable to various volumetric and video medical data, which injects a series of tunable 3D adapters into each transformer block of the image encoder and fine-tunes them together with the mask decoder.
The extensive experiments on 10 datasets demonstrate that MA-SAM consistently outperforms various state-of-the-art 3D approaches without using any prompt and surpasses nnU-Net by 0.9\%, 2.6\%, and 9.9\% in DSC scores for CT multi-organ segmentation, MRI prostate segmentation, and surgical scene segmentation, respectively.
Li~\etal~\cite{li2023promise} propose a prompt-driven 3D medical image segmentation model (ProMISe), which inserts lightweight adapters to extract depth-related spatial context without updating the pre-trained weights for 3D medical image segmentation. 
The evaluation of ProMISe on colon and pancreas tumor segmentation datasets suggests its superior performance compared to the SOTA methods.
Bui~\etal~\cite{bui2023sam3d} introduce SAM3D, which initially applies SAM to process each of the input slices individually, produce slice embeddings, and decode them by a lightweight 3D decoder to ultimately yield the segmentation result.

\subsubsection{Training from Scratch}
In contrast to the works that capture 3D spatial information via 2D to 3D adaptation, Wang~\etal~\cite{SAM-Med3D} propose SAM-Med3D, a volumetric medical image segmentation model with a fully learnable 3D SAM-like architecture.
SAM-Med3D is trained on a large-scale 3D dataset comprising 21K medical images and 131K masks with 247 categories.
The most comprehensive assessment to date is conducted utilizing 15 public datasets, demonstrating its competitive performance with significantly fewer prompt points than the top-performing fine-tuned SAM in the medical domain.
Motivated by the architecture of SAM, Du~\etal~\cite{du2023segvol} propose an interactive volumetric medical image segmentation model named SegVol for CT volume segmentation.
By training on 90k unlabeled CT volumes and 6k labeled ones, SegVol supports the segmentation of over 200 anatomical categories using both spatial and textual prompts, and outperforms the SOTA methods by a large margin on multiple segmentation benchmarks.

\subsection{Summary}
In this section, we review the current research landscape of SAM adaptation for medical image segmentation, which encompasses several distinct but interrelated aspects.
Some studies have delved into fine-tuning strategies, directly tailoring SAM's parameters specifically for medical image segmentation. 
Auto-prompting adaptation approaches explore the automated prompt mechanisms to enhance SAM's flexibility and robustness. 
Framework modification methods initiatives seek to refine SAM's architecture or integration into novel training frameworks, ensuring optimal performance in medical image segmentation scenarios. 
Furthermore, a pivotal exploration is the extension of SAM towards handling 3D medical images, an endeavor initiated to overcome its original limitation of primarily addressing 2D data.
Each of these directions bolsters SAM's effectiveness in addressing the diverse characteristics and complexities inherent in medical image segmentation tasks, outperforming task-specific models across various modalities and targets of interest.

\begin{figure*}
	\includegraphics[width=18cm]{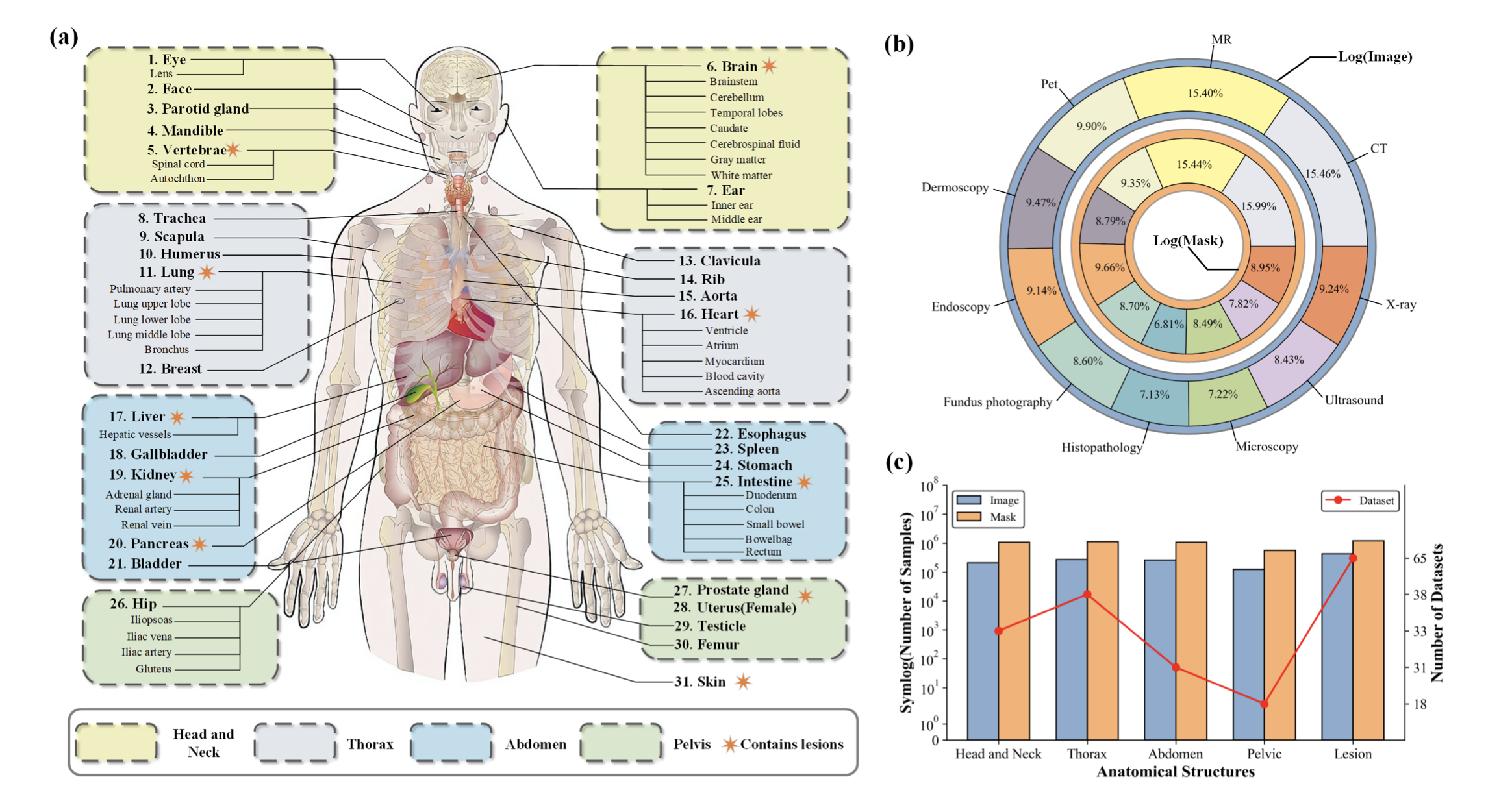}
	\caption{Overview of the SA-Med2D-20M Dataset~\cite{SA-Med2D-20M-Dataset}, a large-scale 2D medical image segmentation dataset featuring a vast collection of 4.6M 2D medical images and 19.7M corresponding masks}
	\label{SA-Med2D-20M-Dataset}
\end{figure*}

\section{Discussion and Conclusion}
In this paper, we provide an overview of recent efforts applying SAM to medical image segmentation tasks. 
When directly employing SAM to medical image segmentation without any adaptation, the performance varies significantly across different datasets and tasks, which indicates SAM's challenge in consistently and accurately achieving zero-shot segmentation on multi-modal and multi-target medical datasets. 
The complexity and diversity of imaging modalities and targets of interest can affect SAM’s segmentation efficacy, particularly manifesting in suboptimal outcomes for objects with irregular shapes, weak boundaries, small sizes, or low contrast.
The subpar segmentation performance of SAM is often insufficient, particularly in medical image segmentation where exceptionally high accuracy is imperative.
To bridge the substantial domain gap between natural images and medical images, several studies~\cite{MedSAM,Med-SAM-Adapter,SAM-Med2D} have explored suitable adaptation strategy that can improve SAM's segmentation results to a certain extent, achieving competitive performance when compared with task-specific models.
While the current performance of SAM may sometimes lack the stability seen in task-specific models, we believe its strong potential to serve as an effective tool for advancing valuable applications in clinical scenarios.
Despite its success, we acknowledge existing challenges and outline potential future directions for enhancement and refinement as follows.

\subsection{Building Large-Scale Medical Datasets}
The evaluation results of several studies~\cite{SAM-MI,SAM-SZU} across various datasets and modalities reveal that directly applying SAM to medical image segmentation does not yield satisfactory performance due to the significant differences between natural images and medical images.
Although fine-tuning SAM on a specific medical dataset can improve the performance, its performance still remains constrained when generalizing to other unseen tasks.

To tackle this issue, it is of great importance to build large-scale medical datasets comprising various modalities and targets of interest for developing universal medical segmentation foundation models. 
Several recent studies have focus on creating large-scale medical datasets~\cite{SAM-SZU,SA-Med2D-20M-Dataset,SAM-Med3D,du2023segvol} by collecting existing public datasets and releasing private datasets.
A notable example is the SA-Med2D-20M Dataset~\cite{SA-Med2D-20M-Dataset}, a recent public large-scale 2D medical image segmentation dataset featuring a vast collection of 4.6M 2D medical images and 19.7M corresponding masks.
This dataset spans the entirety of the human body and offers substantial diversity.
We anticipate that such initiatives will significantly contribute to future advancements in medical foundation models and further propel the research community forward~\cite{moor2023foundation,willemink2022toward}.

\begin{figure*}
	\includegraphics[width=18cm]{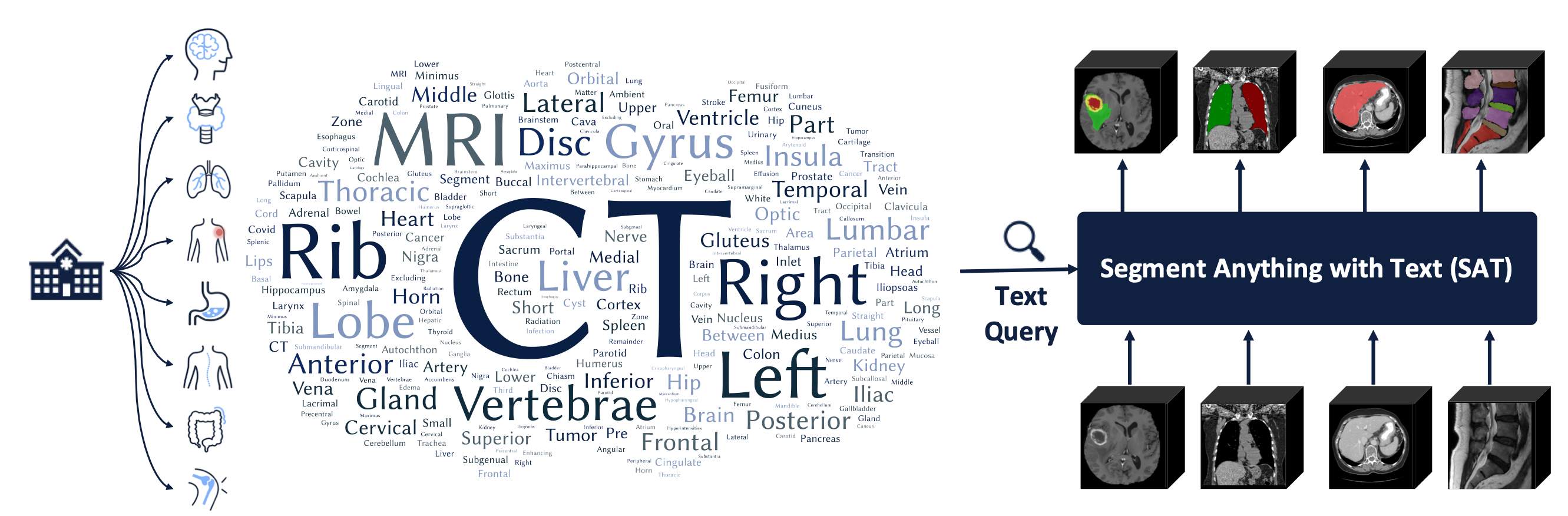}
	\caption{Segment Anything with Text (SAT)~\cite{zhao2023one} utilizes text prompts as queries to perform a wide range of medical image segmentation tasks across different modalities, anatomies, and body regions.}
	\label{SAT}
\end{figure*}

\begin{figure*}
	\includegraphics[width=18cm]{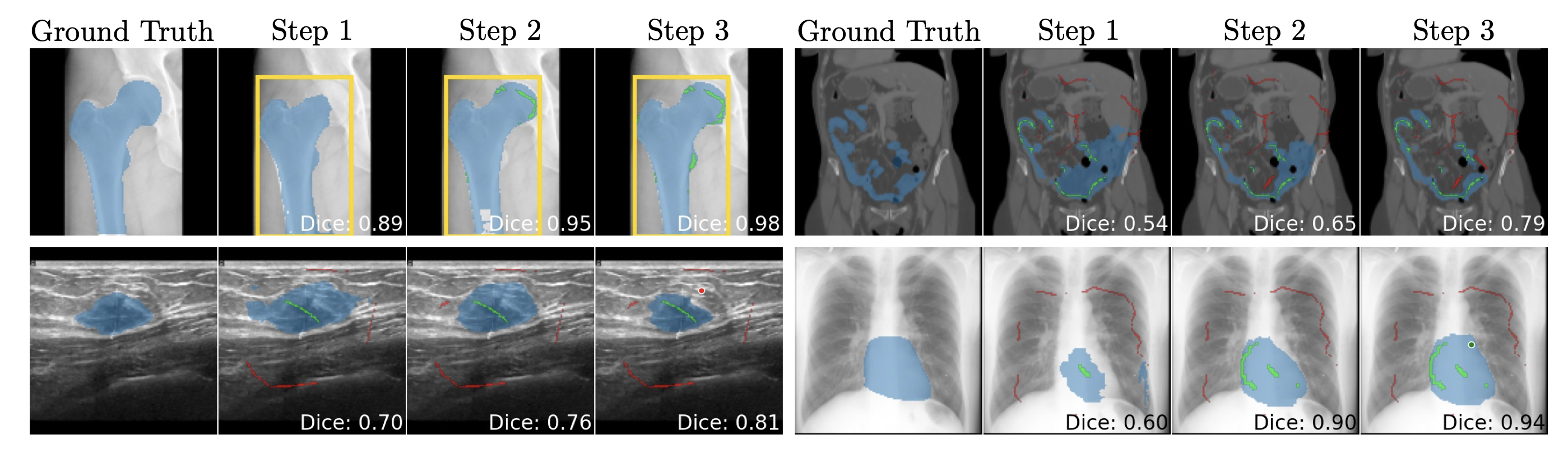}
	\caption{ScribblePrompt~\cite{scribbleprompt} uses scribbles, clicks, and boxes for interactive medical image segmentation. The examples visualize positive scribble and click prompts in \textcolor{green}{green}, negative scribble and click prompts in \textcolor{red}{red}, box prompts in \textcolor{yellow}{yellow}, and the predicted segmentation results in \textcolor{blue}{blue}.}
	\label{Scribble}
\end{figure*}

\subsection{Accelerating Medical Image Annotation}
Building large-scale medical datasets is crucial for the development of medical foundation models, but the heavy annotation costs pose a great challenge.
Developing segmentation models for medical images typically demands domain-specific expertise to provide reliable and accurate annotations~\cite{SemiSurvey,MIA-Imperfect}, which raises expensive annotation costs compared to natural images.
This is particularly evident in the case of commonly used 3D volumetric medical data, where experts must meticulously delineate objects slice by slice, bringing a labor-intensive and time-consuming workload to the annotation procedure.

Although the segmentation results generated by SAM are not always perfect, these segmentation masks can still be considered to accelerate the annotation procedure~\cite{qu2023annotating}.
Instead of labeling targets from scratch, experts can utilize SAM to implement a coarse segmentation and then revise the segmentation manually to achieve fast interactive segmentation. 
This direction is also explored by several recent studies. 
For example, Liu \etal~\cite{SAM-3DSlicer} extend SAM into commonly used medical image viewers with 3D Slicer~\cite{3Dslicer}, which enables researchers to conduct segmentation on medical images with only a latency of 0.6 seconds. The segmentation process through prompt can be automatically applied to the next slice when a slice’s segmentation is complete.
Wang~\etal~\cite{sam-annotation} propose $SAM^{Med}$ framework to leverage SAM for medical image annotation, which consists of two submodules for automatic annotation generation with $SAM^{auto}$ and assisting users in efficiently annotating medical images with $SAM^{assist}$. 
Shen~\etal~\cite{SAM-TEPO} utilize SAM's zero-shot capabilities for interactive medical image segmentation with an innovative reinforcement learning-based framework named temporally-extended prompts optimization (TEPO) by adaptively providing suitable prompt forms for human experts.
Huang~\etal~\cite{huang2023push} propose a label corruption framework to push the boundary of SAM-based segmentation for pseudo-label correction by utilizing a novel noise detection module to distinguish noisy labels from clean labels with uncertainty-based self-correction.
Ning~\etal~\cite{ning2023accurate} utilize a semi-automatic annotation workflow with the assistance of SAM to speed up the annotation process for optical coherence tomography angiography (OCTA).
Ongoing advancements are encouraged to mitigate the annotation costs associated with creating large-scale datasets for a broader variety of medical imaging tasks.

\subsection{Incorporating Scribble and Text Prompts}
Several empirical studies~\cite{SAM-DKFZ-Abdomen,SAM-MI} illustrate that employing box prompts tends to yield superior results compared to point prompts due to the acquisition of relatively more accurate location information.
However, if there are multiple similar instances around the segmentation target, the use of a large bounding box may introduce confusion, potentially resulting in inaccurate segmentation outcomes.
In addition to point and box prompts, the interaction via scribble prompts has become prevalent in medical image segmentation~\cite{ZScribbleseg}, which is useful and efficient when incorporated into SAM. 
Combining scribble prompts with point or box prompts~\cite{scribbleprompt} is an intuitive and effective strategy to tackle non-compact targets with irregular shapes, especially for vessels, intestines, and bones that exhibit shapes characterized by continuity and curvature.
Beyond prompts with positional information, text prompts have also emerged as an intuitive way to inject clinical knowledge into medical image segmentation. 
For example, Zhao~\etal~\cite{zhao2023one} incorporates text prompts into SAM by constructing a multi-modal medical knowledge tree to combine multiple knowledge sources, which can be applied to medical image segmentation tasks across different modalities, anatomies, and body regions.

\subsection{Towards Multi-Modal Medical Images}
Multi-modal medical images play a crucial role in clinical applications due to their ability to provide complementary information about the anatomy, function, and pathology of the human body~\cite{moon2022multi}.
For example, by combining structural MRI with functional MRI (fMRI), clinicians can assess both anatomical structures and functional properties of tissues. 
Combining PET images with corresponding CT or MRI scans allows for the simultaneous assessment of metabolic activity (from PET) and detailed anatomical localization (from CT or MRI). 
This is widely used in oncology to precisely identify and characterize tumors, determine their metabolic activity, assess their relationship to surrounding tissues, and enable more accurate diagnosis and treatment planning.
Extending SAM to learn representations from diverse input modalities can potentially improve generalization across different patient populations and imaging protocols, making them a promising approach for advancing clinical applications.

\subsection{Assisting in More Clinical Applications}
In addition to the existing methodological adaption of SAM for medical image segmentation tasks, some researchers are exploring its integration into more clinical applications dealing with various tasks.
One such application involves GazeSAM~\cite{GazeSAM}, which investigate the potential integration of SAM and eye-tracking technology, designs a collaborative human-computer interaction system, and enables radiologists to acquire segmentation masks by simply looking at the region of interest. 
Ning~\etal~\cite{SAM-UIG} discuss the potential contribution of SAM to enable universal intelligent ultrasound image guidance.
Jiang~\etal~\cite{jiang2023glanceseg} propose a human-in-the-loop, label-free early DR diagnosis framework based on SAM which enables real-time segmentation. 
Song~\etal~\cite{song2023uni} leverage the semantic prior of SAM to supervise the training of a unified framework for MRI cross-modality synthesis and image super-resolution, ensuring the authentic preservation of anatomical structures during synthesis.
SAM's preliminary segmentation capabilities also prove valuable in identifying complex cases that necessitate in-depth scrutiny, thereby alleviating the burden on clinical experts.
Moreover, SAM could aid in minimizing inter-observer variability, which is a prevalent issue in manual contouring~\cite{lappas2022interobserver}.

\subsection{Summary}
Over the past year, we have witnessed unprecedented developments of SAM in medical image segmentation, significantly advancing the development of universal foundation models for medical image analysis.
This comprehensive review aims to provide the community with valuable insights into the trajectory of foundation model development for medical image segmentation.
We anticipate that this reflection will foster a deeper understanding of future directions and inspire further research aimed at creating clinically applicable artificial intelligence.

\section*{Declaration of Competing Interest}
The authors have no conflict of interest to disclose.

\bibliographystyle{IEEEtran}
\bibliography{ref}

\begin{thebibliography}{100}
\providecommand{\url}[1]{#1}
\csname url@samestyle\endcsname
\providecommand{\newblock}{\relax}
\providecommand{\bibinfo}[2]{#2}
\providecommand{\BIBentrySTDinterwordspacing}{\spaceskip=0pt\relax}
\providecommand{\BIBentryALTinterwordstretchfactor}{4}
\providecommand{\BIBentryALTinterwordspacing}{\spaceskip=\fontdimen2\font plus
\BIBentryALTinterwordstretchfactor\fontdimen3\font minus
  \fontdimen4\font\relax}
\providecommand{\BIBforeignlanguage}[2]{{%
\expandafter\ifx\csname l@#1\endcsname\relax
\typeout{** WARNING: IEEEtran.bst: No hyphenation pattern has been}%
\typeout{** loaded for the language `#1'. Using the pattern for}%
\typeout{** the default language instead.}%
\else
\language=\csname l@#1\endcsname
\fi
#2}}
\providecommand{\BIBdecl}{\relax}
\BIBdecl

\bibitem{MIA2017survey}
G.~Litjens, T.~Kooi, B.~E. Bejnordi, A.~A.~A. Setio, F.~Ciompi, M.~Ghafoorian,
  J.~A. Van Der~Laak, B.~Van~Ginneken, and C.~I. S{\'a}nchez, ``A survey on
  deep learning in medical image analysis,'' \emph{Medical image analysis},
  vol.~42, pp. 60--88, 2017.

\bibitem{zhou2021review}
S.~K. Zhou, H.~Greenspan, C.~Davatzikos, J.~S. Duncan, B.~Van~Ginneken,
  A.~Madabhushi, J.~L. Prince, D.~Rueckert, and R.~M. Summers, ``A review of
  deep learning in medical imaging: Imaging traits, technology trends, case
  studies with progress highlights, and future promises,'' \emph{Proceedings of
  the IEEE}, vol. 109, no.~5, pp. 820--838, 2021.

\bibitem{AbdomenCT-1K}
J.~Ma, Y.~Zhang, S.~Gu, C.~Zhu, C.~Ge, Y.~Zhang, X.~An, C.~Wang, Q.~Wang,
  X.~Liu, S.~Cao, Q.~Zhang, S.~Liu, Y.~Wang, Y.~Li, J.~He, and X.~Yang,
  ``Abdomenct-1k: Is abdominal organ segmentation a solved problem?''
  \emph{IEEE Transactions on Pattern Analysis and Machine Intelligence},
  vol.~44, no.~10, pp. 6695--6714, 2022.

\bibitem{medicaldecathlon}
M.~Antonelli, A.~Reinke, S.~Bakas, K.~Farahani, A.~Kopp-Schneider, B.~A.
  Landman, G.~Litjens, B.~Menze, O.~Ronneberger, R.~M. Summers \emph{et~al.},
  ``The medical segmentation decathlon,'' \emph{Nature communications},
  vol.~13, no.~1, p. 4128, 2022.

\bibitem{totalsegmentator}
J.~Wasserthal, H.-C. Breit, M.~T. Meyer, M.~Pradella, D.~Hinck, A.~W. Sauter,
  T.~Heye, D.~T. Boll, J.~Cyriac, S.~Yang \emph{et~al.}, ``Totalsegmentator:
  Robust segmentation of 104 anatomic structures in ct images,''
  \emph{Radiology: Artificial Intelligence}, vol.~5, no.~5, 2023.

\bibitem{wang2023large}
X.~Wang, G.~Chen, G.~Qian, P.~Gao, X.-Y. Wei, Y.~Wang, Y.~Tian, and W.~Gao,
  ``Large-scale multi-modal pre-trained models: A comprehensive survey,''
  \emph{Machine Intelligence Research}, pp. 1--36, 2023.

\bibitem{liang2022foundations}
P.~P. Liang, A.~Zadeh, and L.-P. Morency, ``Foundations and recent trends in
  multimodal machine learning: Principles, challenges, and open questions,''
  \emph{arXiv preprint arXiv:2209.03430}, 2022.

\bibitem{awais2023foundational}
M.~Awais, M.~Naseer, S.~Khan, R.~M. Anwer, H.~Cholakkal, M.~Shah, M.-H. Yang,
  and F.~S. Khan, ``Foundational models defining a new era in vision: A survey
  and outlook,'' \emph{arXiv preprint arXiv:2307.13721}, 2023.

\bibitem{ma2023towards}
J.~Ma and B.~Wang, ``Towards foundation models of biological image
  segmentation,'' \emph{Nature Methods}, vol.~20, no.~7, pp. 953--955, 2023.

\bibitem{SAM-Meta}
A.~Kirillov, E.~Mintun, N.~Ravi, H.~Mao, C.~Rolland, L.~Gustafson, T.~Xiao,
  S.~Whitehead, A.~C. Berg, W.-Y. Lo \emph{et~al.}, ``Segment anything,''
  \emph{arXiv preprint arXiv:2304.02643}, 2023.

\bibitem{SAM-Empirical}
M.~A. Mazurowski, H.~Dong, H.~Gu, J.~Yang, N.~Konz, and Y.~Zhang, ``Segment
  anything model for medical image analysis: an experimental study,''
  \emph{Medical Image Analysis}, vol.~89, p. 102918, 2023.

\bibitem{SAM-SZU}
Y.~Huang, X.~Yang, L.~Liu, H.~Zhou, A.~Chang, X.~Zhou, R.~Chen, J.~Yu, J.~Chen,
  C.~Chen \emph{et~al.}, ``Segment anything model for medical images?''
  \emph{Medical Image Analysis}, p. 103061, 2023.

\bibitem{MedSAM}
J.~Ma and B.~Wang, ``Segment anything in medical images,'' \emph{arXiv preprint
  arXiv:2304.12306}, 2023.

\bibitem{Med-SAM-Adapter}
J.~Wu, R.~Fu, H.~Fang, Y.~Liu, Z.~Wang, Y.~Xu, Y.~Jin, and T.~Arbel, ``Medical
  sam adapter: Adapting segment anything model for medical image
  segmentation,'' \emph{arXiv preprint arXiv:2304.12620}, 2023.

\bibitem{DeSAM}
Y.~Gao, W.~Xia, D.~Hu, and X.~Gao, ``Desam: Decoupling segment anything model
  for generalizable medical image segmentation,'' \emph{arXiv preprint
  arXiv:2306.00499}, 2023.

\bibitem{3DSAM-adapter}
S.~Gong, Y.~Zhong, W.~Ma, J.~Li, Z.~Wang, J.~Zhang, P.-A. Heng, and Q.~Dou,
  ``3dsam-adapter: Holistic adaptation of sam from 2d to 3d for promptable
  medical image segmentation,'' \emph{arXiv preprint arXiv:2306.13465}, 2023.

\bibitem{MedLSAM}
W.~Lei, X.~Wei, X.~Zhang, K.~Li, and S.~Zhang, ``Medlsam: Localize and segment
  anything model for 3d medical images,'' \emph{arXiv preprint
  arXiv:2306.14752}, 2023.

\bibitem{SurgicalSAM}
W.~Yue, J.~Zhang, K.~Hu, Y.~Xia, J.~Luo, and Z.~Wang, ``Surgicalsam: Efficient
  class promptable surgical instrument segmentation,'' \emph{arXiv preprint
  arXiv:2308.08746}, 2023.

\bibitem{SAM-Med2D}
J.~Cheng, J.~Ye, Z.~Deng, J.~Chen, T.~Li, H.~Wang, Y.~Su, Z.~Huang, J.~Chen,
  L.~Jiang \emph{et~al.}, ``Sam-med2d,'' \emph{arXiv preprint
  arXiv:2308.16184}, 2023.

\bibitem{bui2023sam3d}
N.-T. Bui, D.-H. Hoang, M.-T. Tran, and N.~Le, ``Sam3d: Segment anything model
  in volumetric medical images,'' \emph{arXiv preprint arXiv:2309.03493}, 2023.

\bibitem{lin2023samus}
X.~Lin, Y.~Xiang, L.~Zhang, X.~Yang, Z.~Yan, and L.~Yu, ``Samus: Adapting
  segment anything model for clinically-friendly and generalizable ultrasound
  image segmentation,'' \emph{arXiv preprint arXiv:2309.06824}, 2023.

\bibitem{MA-SAM}
C.~Chen, J.~Miao, D.~Wu, Z.~Yan, S.~Kim, J.~Hu, A.~Zhong, Z.~Liu, L.~Sun, X.~Li
  \emph{et~al.}, ``Ma-sam: Modality-agnostic sam adaptation for 3d medical
  image segmentation,'' \emph{arXiv preprint arXiv:2309.08842}, 2023.

\bibitem{li2023nnsam}
Y.~Li, B.~Jing, X.~Feng, Z.~Li, Y.~He, J.~Wang, and Y.~Zhang, ``nnsam:
  Plug-and-play segment anything model improves nnunet performance,''
  \emph{arXiv preprint arXiv:2309.16967}, 2023.

\bibitem{SAM-Med3D}
H.~Wang, S.~Guo, J.~Ye, Z.~Deng, J.~Cheng, T.~Li, J.~Chen, Y.~Su, Z.~Huang,
  Y.~Shen, B.~Fu \emph{et~al.}, ``Sam-med3d,'' \emph{arXiv preprint
  arXiv:2310.15161}, 2023.

\bibitem{li2023promise}
H.~Li, H.~Liu, D.~Hu, J.~Wang, and I.~Oguz, ``Promise: Prompt-driven 3d medical
  image segmentation using pretrained image foundation models,'' \emph{arXiv
  preprint arXiv:2310.19721}, 2023.

\bibitem{du2023segvol}
Y.~Du, F.~Bai, T.~Huang, and B.~Zhao, ``Segvol: Universal and interactive
  volumetric medical image segmentation,'' \emph{arXiv preprint
  arXiv:2311.13385}, 2023.

\bibitem{scribbleprompt}
H.~E. Wong, M.~Rakic, J.~Guttag, and A.~V. Dalca, ``Scribbleprompt: Fast and
  flexible interactive segmentation for any medical image,'' \emph{arXiv
  preprint arXiv:2312.07381}, 2023.

\bibitem{zhang2023semisam}
Y.~Zhang, Y.~Cheng, and Y.~Qi, ``Semisam: Exploring sam for enhancing
  semi-supervised medical image segmentation with extremely limited
  annotations,'' \emph{arXiv preprint arXiv:2312.06316}, 2023.

\bibitem{GPT-3}
T.~Brown, B.~Mann, N.~Ryder, M.~Subbiah, J.~D. Kaplan, P.~Dhariwal,
  A.~Neelakantan, P.~Shyam, G.~Sastry, A.~Askell \emph{et~al.}, ``Language
  models are few-shot learners,'' \emph{Advances in neural information
  processing systems}, vol.~33, pp. 1877--1901, 2020.

\bibitem{GPT-4}
OpenAI, ``Gpt-4 technical report,'' \emph{arXiv preprint arXiv:2303.08774},
  2023.

\bibitem{CLIP}
A.~Radford, J.~W. Kim, C.~Hallacy, A.~Ramesh, G.~Goh, S.~Agarwal, G.~Sastry,
  A.~Askell, P.~Mishkin, J.~Clark \emph{et~al.}, ``Learning transferable visual
  models from natural language supervision,'' in \emph{International conference
  on machine learning}.\hskip 1em plus 0.5em minus 0.4em\relax PMLR, 2021, pp.
  8748--8763.

\bibitem{ALIGN}
C.~Jia, Y.~Yang, Y.~Xia, Y.-T. Chen, Z.~Parekh, H.~Pham, Q.~Le, Y.-H. Sung,
  Z.~Li, and T.~Duerig, ``Scaling up visual and vision-language representation
  learning with noisy text supervision,'' in \emph{International Conference on
  Machine Learning}.\hskip 1em plus 0.5em minus 0.4em\relax PMLR, 2021, pp.
  4904--4916.

\bibitem{DALL-E}
A.~Ramesh, M.~Pavlov, G.~Goh, S.~Gray, C.~Voss, A.~Radford, M.~Chen, and
  I.~Sutskever, ``Zero-shot text-to-image generation,'' in \emph{International
  Conference on Machine Learning}.\hskip 1em plus 0.5em minus 0.4em\relax PMLR,
  2021, pp. 8821--8831.

\bibitem{liu2023summary}
Y.~Liu, T.~Han, S.~Ma, J.~Zhang, Y.~Yang, J.~Tian, H.~He, A.~Li, M.~He, Z.~Liu
  \emph{et~al.}, ``Summary of chatgpt/gpt-4 research and perspective towards
  the future of large language models,'' \emph{arXiv preprint
  arXiv:2304.01852}, 2023.

\bibitem{yi2023towards}
H.~Yi, Z.~Qin, Q.~Lao, W.~Xu, Z.~Jiang, D.~Wang, S.~Zhang, and K.~Li, ``Towards
  general purpose medical ai: Continual learning medical foundation model,''
  \emph{arXiv preprint arXiv:2303.06580}, 2023.

\bibitem{zhang2023foundation}
S.~Zhang and D.~N. Metaxas, ``On the challenges and perspectives of foundation
  models for medical image analysis,'' \emph{arXiv preprint arXiv:2306.05705},
  2023.

\bibitem{li2023artificial}
X.~Li, L.~Zhang, Z.~Wu, Z.~Liu, L.~Zhao, Y.~Yuan, J.~Liu, G.~Li, D.~Zhu, P.~Yan
  \emph{et~al.}, ``Artificial general intelligence for medical imaging,''
  \emph{arXiv preprint arXiv:2306.05480}, 2023.

\bibitem{attention-Nips17}
A.~Vaswani, N.~Shazeer, N.~Parmar, J.~Uszkoreit, L.~Jones, A.~N. Gomez,
  {\L}.~Kaiser, and I.~Polosukhin, ``Attention is all you need,''
  \emph{Advances in neural Information Processing Systems}, vol.~30, 2017.

\bibitem{ViT2020}
A.~Dosovitskiy, L.~Beyer, A.~Kolesnikov, D.~Weissenborn, X.~Zhai,
  T.~Unterthiner, M.~Dehghani, M.~Minderer, G.~Heigold, S.~Gelly \emph{et~al.},
  ``An image is worth 16x16 words: Transformers for image recognition at
  scale,'' in \emph{International Conference on Learning Representations},
  2020.

\bibitem{MAE}
K.~He, X.~Chen, S.~Xie, Y.~Li, P.~Doll{\'a}r, and R.~Girshick, ``Masked
  autoencoders are scalable vision learners,'' in \emph{Proceedings of the
  IEEE/CVF Conference on Computer Vision and Pattern Recognition}, 2022, pp.
  16\,000--16\,009.

\bibitem{FourierPE-Nips20}
M.~Tancik, P.~Srinivasan, B.~Mildenhall, S.~Fridovich-Keil, N.~Raghavan,
  U.~Singhal, R.~Ramamoorthi, J.~Barron, and R.~Ng, ``Fourier features let
  networks learn high frequency functions in low dimensional domains,''
  \emph{Advances in Neural Information Processing Systems}, vol.~33, pp.
  7537--7547, 2020.

\bibitem{Lin2017FocalLF}
T.-Y. Lin, P.~Goyal, R.~B. Girshick, K.~He, and P.~Doll{\'a}r, ``Focal loss for
  dense object detection,'' \emph{2017 IEEE International Conference on
  Computer Vision (ICCV)}, pp. 2999--3007, 2017.

\bibitem{Milletar2016VNetFC}
F.~Milletar{\`i}, N.~Navab, and S.-A. Ahmadi, ``V-net: Fully convolutional
  neural networks for volumetric medical image segmentation,'' \emph{2016
  Fourth International Conference on 3D Vision (3DV)}, pp. 565--571, 2016.

\bibitem{SAM-DKFZ-Abdomen}
S.~Roy, T.~Wald, G.~Koehler, M.~R. Rokuss, N.~Disch, J.~Holzschuh, D.~Zimmerer,
  and K.~H. Maier-Hein, ``Sam. md: Zero-shot medical image segmentation
  capabilities of the segment anything model,'' \emph{arXiv preprint
  arXiv:2304.05396}, 2023.

\bibitem{ji2022amos}
Y.~Ji, H.~Bai, C.~Ge, J.~Yang, Y.~Zhu, R.~Zhang, Z.~Li, L.~Zhanng, W.~Ma,
  X.~Wan \emph{et~al.}, ``Amos: A large-scale abdominal multi-organ benchmark
  for versatile medical image segmentation,'' \emph{Advances in Neural
  Information Processing Systems}, vol.~35, pp. 36\,722--36\,732, 2022.

\bibitem{SAM-LiverTumor}
C.~Hu and X.~Li, ``When sam meets medical images: An investigation of segment
  anything model (sam) on multi-phase liver tumor segmentation,'' \emph{arXiv
  preprint arXiv:2304.08506}, 2023.

\bibitem{U-Net}
O.~Ronneberger, P.~Fischer, and T.~Brox, ``U-net: Convolutional networks for
  biomedical image segmentation,'' in \emph{International Conference on Medical
  image computing and computer-assisted intervention}.\hskip 1em plus 0.5em
  minus 0.4em\relax Springer, 2015, pp. 234--241.

\bibitem{SAM-BrainMR}
S.~Mohapatra, A.~Gosai, and G.~Schlaug, ``Sam vs bet: A comparative study for
  brain extraction and segmentation of magnetic resonance images using deep
  learning,'' \emph{arXiv preprint arXiv:2304.04738}, 2023.

\bibitem{SAM-BraTS}
P.~Zhang and Y.~Wang, ``Segment anything model for brain tumor segmentation,''
  \emph{arXiv preprint arXiv:2309.08434}, 2023.

\bibitem{SAM-pathology}
R.~Deng, C.~Cui, Q.~Liu, T.~Yao, L.~W. Remedios, S.~Bao, B.~A. Landman, L.~E.
  Wheless, L.~A. Coburn, K.~T. Wilson \emph{et~al.}, ``Segment anything model
  (sam) for digital pathology: Assess zero-shot segmentation on whole slide
  imaging,'' \emph{arXiv preprint arXiv:2304.04155}, 2023.

\bibitem{SAM-Polyps}
T.~Zhou, Y.~Zhang, Y.~Zhou, Y.~Wu, and C.~Gong, ``Can sam segment polyps?''
  \emph{arXiv preprint arXiv:2304.07583}, 2023.

\bibitem{silva2014toward}
J.~Silva, A.~Histace, O.~Romain, X.~Dray, and B.~Granado, ``Toward embedded
  detection of polyps in wce images for early diagnosis of colorectal cancer,''
  \emph{International journal of computer assisted radiology and surgery},
  vol.~9, pp. 283--293, 2014.

\bibitem{bernal2015wm}
J.~Bernal, F.~J. S{\'a}nchez, G.~Fern{\'a}ndez-Esparrach, D.~Gil,
  C.~Rodr{\'\i}guez, and F.~Vilari{\~n}o, ``Wm-dova maps for accurate polyp
  highlighting in colonoscopy: Validation vs. saliency maps from physicians,''
  \emph{Computerized medical imaging and graphics}, vol.~43, pp. 99--111, 2015.

\bibitem{tajbakhsh2015automated}
N.~Tajbakhsh, S.~R. Gurudu, and J.~Liang, ``Automated polyp detection in
  colonoscopy videos using shape and context information,'' \emph{IEEE
  transactions on medical imaging}, vol.~35, no.~2, pp. 630--644, 2015.

\bibitem{vazquez2017benchmark}
D.~V{\'a}zquez, J.~Bernal, F.~J. S{\'a}nchez, G.~Fern{\'a}ndez-Esparrach, A.~M.
  L{\'o}pez, A.~Romero, M.~Drozdzal, A.~Courville \emph{et~al.}, ``A benchmark
  for endoluminal scene segmentation of colonoscopy images,'' \emph{Journal of
  healthcare engineering}, vol. 2017, 2017.

\bibitem{jha2020kvasir}
D.~Jha, P.~H. Smedsrud, M.~A. Riegler, P.~Halvorsen, T.~de~Lange, D.~Johansen,
  and H.~D. Johansen, ``Kvasir-seg: A segmented polyp dataset,'' in
  \emph{MultiMedia Modeling: 26th International Conference, MMM 2020, Daejeon,
  South Korea, January 5--8, 2020, Proceedings, Part II 26}.\hskip 1em plus
  0.5em minus 0.4em\relax Springer, 2020, pp. 451--462.

\bibitem{SAM-RS}
A.-C. Wang, M.~Islam, M.~Xu, Y.~Zhang, and H.~Ren, ``Sam meets robotic surgery:
  An empirical study in robustness perspective,'' \emph{arXiv preprint
  arXiv:2304.14674}, 2023.

\bibitem{allan20192017}
M.~Allan, A.~Shvets, T.~Kurmann, Z.~Zhang, R.~Duggal, Y.-H. Su, N.~Rieke,
  I.~Laina, N.~Kalavakonda, S.~Bodenstedt \emph{et~al.}, ``2017 robotic
  instrument segmentation challenge,'' \emph{arXiv preprint arXiv:1902.06426},
  2019.

\bibitem{allan20202018}
M.~Allan, S.~Kondo, S.~Bodenstedt, S.~Leger, R.~Kadkhodamohammadi, I.~Luengo,
  F.~Fuentes, E.~Flouty, A.~Mohammed, M.~Pedersen \emph{et~al.}, ``2018 robotic
  scene segmentation challenge,'' \emph{arXiv preprint arXiv:2001.11190}, 2020.

\bibitem{SAM-Meds}
S.~He, R.~Bao, J.~Li, P.~E. Grant, and Y.~Ou, ``Accuracy of segment-anything
  model (sam) in medical image segmentation tasks,'' \emph{arXiv preprint
  arXiv:2304.09324}, 2023.

\bibitem{SAM-MI}
D.~Cheng, Z.~Qin, Z.~Jiang, S.~Zhang, Q.~Lao, and K.~Li, ``Sam on medical
  images: A comprehensive study on three prompt modes,'' \emph{arXiv preprint
  arXiv:2305.00035}, 2023.

\bibitem{SAM-RO}
L.~Zhang, Z.~Liu, L.~Zhang, Z.~Wu, X.~Yu, J.~Holmes, H.~Feng, H.~Dai, X.~Li,
  Q.~Li \emph{et~al.}, ``Segment anything model (sam) for radiation oncology,''
  \emph{arXiv preprint arXiv:2306.11730}, 2023.

\bibitem{SAM-ConcealedScenes}
G.-P. Ji, D.-P. Fan, P.~Xu, M.-M. Cheng, B.~Zhou, and L.~V. Gool, ``Sam
  struggles in concealed scenes - empirical study on "segment anything",''
  \emph{arXiv preprint arXiv:2304.06022}, 2023.

\bibitem{SAM-Realworld}
W.~Ji, J.~Li, Q.~Bi, W.~Li, and L.~Cheng, ``Segment anything is not always
  perfect: An investigation of sam on different real-world applications,''
  \emph{arXiv preprint arXiv:2304.05750}, 2023.

\bibitem{skinSAM}
M.~Hu, Y.~Li, and X.~Yang, ``Skinsam: Empowering skin cancer segmentation with
  segment anything model,'' \emph{arXiv preprint arXiv:2304.13973}, 2023.

\bibitem{PolypSAM}
Y.~Li, M.~Hu, and X.~Yang, ``Polyp-sam: Transfer sam for polyp segmentation,''
  \emph{arXiv preprint arXiv:2305.00293}, 2023.

\bibitem{Lora}
E.~J. Hu, Y.~Shen, P.~Wallis, Z.~Allen-Zhu, Y.~Li, S.~Wang, L.~Wang, and
  W.~Chen, ``Lora: Low-rank adaptation of large language models,'' \emph{arXiv
  preprint arXiv:2106.09685}, 2021.

\bibitem{SAMed}
K.~Zhang and D.~Liu, ``Customized segment anything model for medical image
  segmentation,'' \emph{arXiv preprint arXiv:2304.13785}, 2023.

\bibitem{feng2023cheap}
W.~Feng, L.~Zhu, and L.~Yu, ``Cheap lunch for medical image segmentation by
  fine-tuning sam on few exemplars,'' \emph{arXiv preprint arXiv:2308.14133},
  2023.

\bibitem{paranjape2023adaptivesam}
J.~N. Paranjape, N.~G. Nair, S.~Sikder, S.~S. Vedula, and V.~M. Patel,
  ``Adaptivesam: Towards efficient tuning of sam for surgical scene
  segmentation,'' \emph{arXiv preprint arXiv:2308.03726}, 2023.

\bibitem{pandey2023comprehensive}
S.~Pandey, K.-F. Chen, and E.~B. Dam, ``Comprehensive multimodal segmentation
  in medical imaging: Combining yolov8 with sam and hq-sam models,'' in
  \emph{Proceedings of the IEEE/CVF International Conference on Computer
  Vision}, 2023, pp. 2592--2598.

\bibitem{anand2023one}
D.~Anand, V.~Singhal, D.~D. Shanbhag, S.~KS, U.~Patil, C.~Bhushan, K.~Manickam,
  D.~Gui, R.~Mullick, A.~Gopal \emph{et~al.}, ``One-shot localization and
  segmentation of medical images with foundation models,'' \emph{arXiv preprint
  arXiv:2310.18642}, 2023.

\bibitem{AutoSAM}
T.~Shaharabany, A.~Dahan, R.~Giryes, and L.~Wolf, ``Autosam: Adapting sam to
  medical images by overloading the prompt encoder,'' \emph{arXiv preprint
  arXiv:2306.06370}, 2023.

\bibitem{All-in-SAM}
C.~Cui, R.~Deng, Q.~Liu, T.~Yao, S.~Bao, L.~W. Remedios, Y.~Tang, and Y.~Huo,
  ``All-in-sam: from weak annotation to pixel-wise nuclei segmentation with
  prompt-based finetuning,'' \emph{arXiv preprint arXiv:2307.00290}, 2023.

\bibitem{chen2023sam}
T.~Chen, L.~Zhu, C.~Ding, R.~Cao, S.~Zhang, Y.~Wang, Z.~Li, L.~Sun, P.~Mao, and
  Y.~Zang, ``Sam fails to segment anything?--sam-adapter: Adapting sam in
  underperformed scenes: Camouflage, shadow, and more,'' \emph{arXiv preprint
  arXiv:2304.09148}, 2023.

\bibitem{xu2023eviprompt}
Y.~Xu, J.~Tang, A.~Men, and Q.~Chen, ``Eviprompt: A training-free evidential
  prompt generation method for segment anything model in medical images,''
  \emph{arXiv preprint arXiv:2311.06400}, 2023.

\bibitem{SAM-U}
G.~Deng, K.~Zou, K.~Ren, M.~Wang, X.~Yuan, S.~Ying, and H.~Fu, ``Sam-u:
  Multi-box prompts triggered uncertainty estimation for reliable sam in
  medical image,'' \emph{arXiv preprint arXiv:2307.04973}, 2023.

\bibitem{UR-SAM}
Y.~Zhang, S.~Hu, C.~Jiang, Y.~Cheng, and Y.~Qi, ``Segment anything model with
  uncertainty rectification for auto-prompting medical image segmentation,''
  \emph{arXiv preprint arXiv:2311.10529}, 2023.

\bibitem{SAM-Path}
J.~Zhang, K.~Ma, S.~Kapse, J.~Saltz, M.~Vakalopoulou, P.~Prasanna, and
  D.~Samaras, ``Sam-path: A segment anything model for semantic segmentation in
  digital pathology,'' \emph{arXiv preprint arXiv:2307.09570}, 2023.

\bibitem{chai2023ladder}
S.~Chai, R.~K. Jain, S.~Teng, J.~Liu, Y.~Li, T.~Tateyama, and Y.-w. Chen,
  ``Ladder fine-tuning approach for sam integrating complementary network,''
  \emph{arXiv preprint arXiv:2306.12737}, 2023.

\bibitem{IA-SAM}
Y.~Zhang, T.~Zhou, P.~Liang, and D.~Z. Chen, ``Input augmentation with sam:
  Boosting medical image segmentation with segmentation foundation model,''
  \emph{arXiv preprint arXiv:2304.11332}, 2023.

\bibitem{zhang2023samdsk}
Y.~Zhang, T.~Zhou, S.~Wang, Y.~Wu, P.~Gu, and D.~Z. Chen, ``Samdsk: Combining
  segment anything model with domain-specific knowledge for semi-supervised
  learning in medical image segmentation,'' \emph{arXiv preprint
  arXiv:2308.13759}, 2023.

\bibitem{li2023segment}
N.~Li, L.~Xiong, W.~Qiu, Y.~Pan, Y.~Luo, and Y.~Zhang, ``Segment anything model
  for semi-supervised medical image segmentation via selecting reliable
  pseudo-labels,'' in \emph{International Conference on Neural Information
  Processing}, 2023.

\bibitem{li2023leverage}
X.~Li, R.~Deng, Y.~Tang, S.~Bao, H.~Yang, and Y.~Huo, ``Leverage weakly
  annotation to pixel-wise annotation via zero-shot segment anything model for
  molecular-empowered learning,'' \emph{arXiv preprint arXiv:2308.05785}, 2023.

\bibitem{2-5D}
Y.~Zhang, Q.~Liao, L.~Ding, and J.~Zhang, ``Bridging 2d and 3d segmentation
  networks for computation-efficient volumetric medical image segmentation: An
  empirical study of 2.5 d solutions,'' \emph{Computerized Medical Imaging and
  Graphics}, p. 102088, 2022.

\bibitem{isensee2021nnu}
F.~Isensee, P.~F. Jaeger, S.~A. Kohl, J.~Petersen, and K.~H. Maier-Hein,
  ``nnu-net: a self-configuring method for deep learning-based biomedical image
  segmentation,'' \emph{Nature methods}, vol.~18, no.~2, pp. 203--211, 2021.

\bibitem{SA-Med2D-20M-Dataset}
J.~Ye, J.~Cheng, J.~Chen, Z.~Deng, T.~Li, H.~Wang, Y.~Su, Z.~Huang, J.~Chen,
  L.~Jiang \emph{et~al.}, ``Sa-med2d-20m dataset: Segment anything in 2d
  medical imaging with 20 million masks,'' \emph{arXiv preprint
  arXiv:2311.11969}, 2023.

\bibitem{moor2023foundation}
M.~Moor, O.~Banerjee, Z.~F.~H. Abad, H.~M. Krumholz, J.~Leskovec, E.~J. Topol,
  and P.~Rajpurkar, ``Foundation models for generalist medical artificial
  intelligence,'' \emph{Nature}, vol. 616, pp. 259--265, 2023.

\bibitem{willemink2022toward}
M.~J. Willemink, H.~R. Roth, and V.~Sandfort, ``Toward foundational deep
  learning models for medical imaging in the new era of transformer networks.''
  \emph{Radiology. Artificial intelligence}, vol.~46, p. 210284, 2022.

\bibitem{zhao2023one}
Z.~Zhao, Y.~Zhang, C.~Wu, X.~Zhang, Y.~Zhang, Y.~Wang, and W.~Xie, ``One model
  to rule them all: Towards universal segmentation for medical images with text
  prompts,'' \emph{arXiv preprint arXiv:2312.17183}, 2023.

\bibitem{SemiSurvey}
R.~Jiao, Y.~Zhang, L.~Ding, R.~Cai, and J.~Zhang, ``Learning with limited
  annotations: A survey on deep semi-supervised learning for medical image
  segmentation,'' \emph{arXiv preprint arXiv:2207.14191}, 2022.

\bibitem{MIA-Imperfect}
N.~Tajbakhsh, L.~Jeyaseelan, Q.~Li, J.~N. Chiang, Z.~Wu, and X.~Ding,
  ``Embracing imperfect datasets: A review of deep learning solutions for
  medical image segmentation,'' \emph{Medical Image Analysis}, vol.~63, p.
  101693, 2020.

\bibitem{qu2023annotating}
C.~Qu, T.~Zhang, H.~Qiao, J.~Liu, Y.~Tang, A.~L. Yuille, and Z.~Zhou, ``Segment
  anything,'' \emph{arXiv preprint arXiv:2305.09666}, 2023.

\bibitem{SAM-3DSlicer}
Y.~Liu, J.~Zhang, Z.~She, A.~Kheradmand, and M.~Armand, ``Samm (segment any
  medical model): A 3d slicer integration to sam,'' \emph{arXiv preprint
  arXiv:2304.05622}, 2023.

\bibitem{3Dslicer}
A.~Fedorov, R.~R. Beichel, J.~Kalpathy-Cramer, J.~Finet, J.-C. Fillion-Robin,
  S.~Pujol, C.~Bauer, D.~L. Jennings, F.~M. Fennessy, M.~Sonka, J.~M. Buatti,
  S.~R. Aylward, J.~V. Miller, S.~D. Pieper, and R.~Kikinis, ``3d slicer as an
  image computing platform for the quantitative imaging network.''
  \emph{Magnetic resonance imaging}, vol. 30 9, pp. 1323--41, 2012.

\bibitem{sam-annotation}
C.~Wang, D.~Li, S.~Wang, C.~Zhang, Y.~Wang, Y.~Liu, and G.~Yang, ``$sam^{Med}$:
  A medical image annotation framework based on large vision model,''
  \emph{arXiv preprint arXiv:2307.05617}, 2023.

\bibitem{SAM-TEPO}
C.~Shen, W.~Li, Y.~Zhang, and X.~Wang, ``Temporally-extended prompts
  optimization for sam in interactive medical image segmentation,'' \emph{arXiv
  preprint arXiv:2306.08958}, 2023.

\bibitem{huang2023push}
Z.~Huang, H.~Liu, H.~Zhang, F.~Xing, A.~Laine, E.~Angelini, C.~Hendon, and
  Y.~Gan, ``Push the boundary of sam: A pseudo-label correction framework for
  medical segmentation,'' \emph{arXiv preprint arXiv:2308.00883}, 2023.

\bibitem{ning2023accurate}
H.~Ning, C.~Wang, X.~Chen, and S.~Li, ``An accurate and efficient neural
  network for octa vessel segmentation and a new dataset,'' \emph{arXiv
  preprint arXiv:2309.09483}, 2023.

\bibitem{ZScribbleseg}
K.~Zhang and X.~Zhuang, ``Zscribbleseg: Zen and the art of scribble supervised
  medical image segmentation,'' \emph{arXiv preprint arXiv:2301.04882}, 2023.

\bibitem{moon2022multi}
J.~H. Moon, H.~Lee, W.~Shin, Y.-H. Kim, and E.~Choi, ``Multi-modal
  understanding and generation for medical images and text via vision-language
  pre-training,'' \emph{IEEE Journal of Biomedical and Health Informatics},
  vol.~26, no.~12, pp. 6070--6080, 2022.

\bibitem{GazeSAM}
B.~Wang, A.~Aboah, Z.~Zhang, and U.~Bagci, ``Gazesam: What you see is what you
  segment,'' \emph{arXiv preprint arXiv:2304.13844}, 2023.

\bibitem{SAM-UIG}
G.~Ning, H.~Liang, Z.~Jiang, H.~Zhang, and H.~Liao, ``The potential of 'segment
  anything' (sam) for universal intelligent ultrasound image guidance.''
  \emph{Bioscience trends}, 2023.

\bibitem{jiang2023glanceseg}
H.~Jiang, M.~Gao, Z.~Liu, C.~Tang, X.~Zhang, S.~Jiang, W.~Yuan, and J.~Liu,
  ``Glanceseg: Real-time microaneurysm lesion segmentation with gaze-map-guided
  foundation model for early detection of diabetic retinopathy,'' \emph{arXiv
  preprint arXiv:2311.08075}, 2023.

\bibitem{song2023uni}
Z.~Song, Z.~Qi, X.~Wang, X.~Zhao, Z.~Shen, S.~Wang, M.~Fei, Z.~Wang, D.~Zang,
  D.~Chen \emph{et~al.}, ``Uni-coal: A unified framework for cross-modality
  synthesis and super-resolution of mr images,'' \emph{arXiv preprint
  arXiv:2311.08225}, 2023.

\bibitem{lappas2022interobserver}
G.~Lappas, N.~Staut, N.~G. Lieuwes, R.~Biemans, C.~J. Wolfs, S.~J. van Hoof,
  L.~J. Dubois, and F.~Verhaegen, ``Inter-observer variability of organ
  contouring for preclinical studies with cone beam computed tomography
  imaging,'' \emph{Physics and Imaging in Radiation Oncology}, vol.~21, pp. 11
  -- 17, 2022.

\end{thebibliography}




%








\end{document}